\documentclass[epj,nopacs,twocolumn]{svjour}

\usepackage[dvips]{color}
\usepackage{fancyhdr}

\usepackage{enumerate}

\usepackage{amsmath}

\usepackage{amssymb}

\usepackage{mathrsfs,bbm}
\usepackage{slashed}
\usepackage{array}
\usepackage{latexsym}

\usepackage{multirow}

\usepackage{color}
\begin{document}
\title{Scalar Potential Without Cubic Term in 3-3-1 Models Without Exotic Electric Charges}
\author{Yithsbey Giraldo\inst{1,2} \and William A. Ponce\inst{2}
}                 

\institute{ Departamento de F\'\i sica, Universidad de Nari\~no,
 A.A. 1175, Pasto, Colombia. \and  Instituto de F\'\i sica, Universidad de Antioquia,
A.A. 1226, Medell\'\i n, Colombia.}
\date{Received: date / Revised version: date}
%
\abstract{
A detailed study of the criteria for stability of the scalar potential, and the proper electroweak symmetry breaking pattern in some 3-3-1 models without exotic electric charges is presented. In this paper we concentrate in a scalar sector with three Higgs scalar triplets, with a potential that does not include the cubic term, due to the presence of a discrete symmetry. For the analysis we use, and improve, a method previously developed to study the scalar potential in the two-Higgs-doublet extension of the standard model. Our main result is to show  the consistency of those 3-3-1 models without exotic electric charges.
\PACS{
      {PACS-key}{discribing text of that key}   \and
      {PACS-key}{discribing text of that key}
     } 
} 
\authorrunning{Yithsbey Giraldo et al.}
\maketitle
\section{Introduction}
\label{sec:sec1}
A simple extension of the standard model~(SM) consists of adding to the model a second Higgs scalar doublet~\cite{dona}, defining in this way the so-called two Higgs doublet model (THDM). The different ways how the two Higgs scalar doublets couple to the fermion sector, define the different versions of this extension\cite{dona,b5}. Many gauge group extensions of the SM have the THDM as an effective low energy theory (in this regard see the papers in \cite{b5} and references therein).
In these extensions one intermediate step in the symmetry breaking chain leads to the $SU(3)_c\otimes SU(2)_L\otimes U(1)_Y$ gauge theory with two Higss doublets in one of its several versions.

A novel method for a detailed analysis of the scalar potential in the most general THDM  was presented in Refs.~\cite{b1,b2} where by using  powerful algebraic techniques, the authors studied in detail the stationary points of the scalar potential. This allowed them to give, in a very concise way, clear criteria for the stability of the scalar potential and for the correct electroweak symmetry breaking pattern. By using different approaches, the authors in Refs.~\cite{n5} reached also interesting conclusions for the scalar potential of the THDM, some of them related to the ones presented in Ref.\cite{b1}.

In the present work we make use of some of the new algebraic developments cited in the former paragraph, to analyze the scalar sector of an extension to the SM based on the local gauge group $SU(3)_c\otimes SU(3)_L\otimes U(1)_X$~\cite{pf,vl,ozer,sher,pfs} (called hereafter 3-3-1 for short).

In general, the scalar sector for 3-3-1 models is quite complicated and difficult to analyze in detail. For example, for the minimal model (the Pisano-Pleitez-Frampton model~\cite{pf}) three Higgs scalar triplets and one additional Higgs sextet must be used, in order to break the symmetry and provide, at the same time, masses to the fermion fields. For the 3-3-1 models without exotic electric charges \cite{vl,ozer,sher,pfs} the situation is simpler because it turns out that less Higgs scalar multiplets are needed~\cite{rob}. For example, the so called economical 3-3-1 model~\cite{b3} makes use of only two  Higgs scalar triplets which are able to break the symmetry in a consistent way, although they are not able to produce a consistent fermion mass spectrum at tree-level. The alternative approach is to deal with three Higgs scalar triplets instead of two, as done for example in Refs.~\cite{vl,ozer}.

In this work we pursue the study of the scalar sector of the 3-3-1 models without exotic electric charges started in Ref.~\cite{yith}, by considering this time a model with three Higgs scalar triplets.  A discrete symmetry will be applied to the corresponding scalar potential~\cite{folo,bb3_1}, which simplifies and facilitates its analysis, due that the cubic (or trilinear) term would be absent. In this analysis we will derive constraints on the parameters of the scalar potential coming from its stability and from the electroweak symmetry breaking conditions (the stability of an scalar potential at the classical level, which is fulfilled when it is bounded from below, is a necessary condition in order to have a consistent theory). The global minimum of the potential will also be found by determining its stationary points.

This paper is organized as follows: in Sect.~\ref{sec:sec2} we introduce the 3-3-1 models without exotic electric charges and review the different scalar sectors available in the literature for this type of models; in Sect.~\ref{sec:sec3} we introduce the scalar potential under study and analyze the consistency of the electroweak symmetry breaking pattern proposed; then in Sect.~\ref{sec:sec4} we study the stability of the scalar potential, and in Sect.~\ref{sec:sec5} we find its stationary points and its global minimum. Our conclusions are presented in Sect.~\ref{sec:sec6}. Three appendixes, one with a general review and two technical ones are presented at the end.
\section{\label{sec:sec2} 3-3-1 models without exotic electric charges}
As demonstrated in Refs.~\cite{sher,pfs,b3}, there exist a total of eight different three-family 3-3-1 models without exotic electric charges, that do not contain fermion singlets or in vector-like representations. Each model has a different spin 1/2 particle structure, but they have the same gauge boson sector. In principle, all can bear the same scalar sector too.
\subsection{\label{sec:sec21} The Gauge sector}
The gauge boson structure for any 3-3-1 model without exotic electric charges is: 
one gauge field $B^\mu$ associated with $U(1)_X$, the 8
gluon fields $G^\mu$ associated with $SU(3)_c$ which remain massless after
breaking the symmetry, and 8 gauge fields from $SU(3)_L$ that we write as \cite{b3}

\[\frac{1}{\sqrt{2}}\lambda_\alpha A^\mu_\alpha=
\frac{1}{\sqrt{2}}
\left(\begin{array}{ccc}D^\mu_1 & W^{+\mu} & K^{+\mu} \\ W^{-\mu} & D^\mu_2 &
K^{0\mu} \\ K^{-\mu} & \bar{K}^{0\mu} & D^\mu_3 \end{array}\right), \]
where $D^\mu_1=A_3^\mu/\sqrt{2}+A_8^\mu/\sqrt{6},\;
D^\mu_2=-A_3^\mu/\sqrt{2}+A_8^\mu/\sqrt{6}$, and
$D^\mu_3=-2A_8^\mu/\sqrt{6}$. $\lambda_i, \; i=1,2,...,8$, are the eight
Gell-Mann matrices normalized as $Tr(\lambda_i\lambda_j)  
=2\delta_{ij}$.

The charge operator associated with the unbroken gauge symmetry $U(1)_Q$ is 
\begin{equation}\label{qele}
Q=\frac{\lambda_{3}}{2}+\frac{\lambda_{8}}{2\sqrt{3}}+XI_3
\end{equation}
where $I_3=Diag.(1,1,1)$ is the diagonal $3\times 3$ unit matrix, and the 
$X$ values are related to the $U(1)_X$ hypercharge and are fixed by 
anomaly cancellation. 
The sine square of the electroweak mixing angle is given by 
$S_W^2=3g_1^2/(3g_3^2+4g_1^2)$ where $g_1$ and $g_3$ are the coupling 
constants of $U(1)_X$ and $SU(3)_L$ respectively. The photon field is 
given by
\begin{equation}\label{foton}
A_0^\mu=S_WA_3^\mu+C_W\left[\frac{T_W}{\sqrt{3}}A_8^\mu + 
\sqrt{(1-T_W^2/3)}B^\mu\right],
\end{equation}
where $C_W$ and $T_W$ are the cosine and tangent of the electroweak mixing 
angle, respectively.

There are two weak neutral currents in the model 
associated with the two flavor diagonal neutral gauge weak bosons 
\begin{eqnarray}\nonumber \label{zzs}
Z_0^\mu&=&C_WA_3^\mu-S_W\left[\frac{T_W}{\sqrt{3}}A_8^\mu + 
\sqrt{(1-T_W^2/3)}B^\mu\right], \\ \label{zetas}
Z_0^{\prime\mu}&=&-\sqrt{(1-T_W^2/3)}A_8^\mu+\frac{T_W}{\sqrt{3}}B^\mu,
\end{eqnarray}
and one current associated with the flavor non diagonal neutral gauge 
boson $K^{0\mu}$ which is charged in the sense that it has a kind of 
weak V isospin charge. In the former expressions 
$Z^\mu_0$ coincides with the weak neutral current of the SM. 
\subsection{\label{sec:sec22} A Fermion sector}
The particular 3-3-1 model without exotic electric charges most extensively studied in the literature, known as the 3-3-1 model with right-handed neutrinos, has the following anomaly free fermion sector~\cite{vl}:
\begin{eqnarray*}
\psi_L^{a}&=&(l^{-a},\nu^a,N^{0a})^T_L\sim (1,3^*,-1/3),\\
l^{+a}_L &\sim& (1,1,1),\\
Q_L^{i}&=&(u^i,d^i,D^i)^T_L\sim (3,3,0),\\
Q_L^{3}&=&(d^3,u^3,U)^T_L\sim (3,3^*,1/3),\\
u^{ca}_L&\sim&(3^*,1,-2/3),\;\; d^{ca}_L\sim (3^*,1,1/3),\\
U^c_L&\sim&(3^*,1,-2/3),\;\; D^{ci}_L\sim (3^*,1,1/3),
\end{eqnarray*}
where the numbers inside the parenthesis stand for $[SU(3)_c,\linebreak SU(3)_L, U(1)_Y]$ representations, $a=1,2,3$ is a family index and $i=1,2$ is related to two of the three families. $D^i$ and $U$ are three exotic quarks with electric charges $-1/3, \; -1/3$ and $2/3$, respectively, and $N^{0a}_L$ can play the role  of right-handed neutrinos. The three lepton families are arranged in antitriplets of $SU(3)_L$. In order to cancel the $SU(3)_L$ anomaly, two quark families must transform as $SU(3)_L$ triplets and the remaining one as an antitriplet. It is customary to arrange the first two quark families in triplets and the third one in an antitriplet. This choice is meant to distinguish the possible new dynamics effects arising in the third family.
\subsection{\label{sec:sec23}The scalar sector}
As far as we know, for the 3-3-1 models without exotic electric charges, three different scalar sectors have been used in the literature,  to deal with the spontaneous breaking of the gauge symmetry down to $U(1)_{Q}$ and, to produce at the same time, masses for the Fermion fields. Each set, as described anon, has its own advantages and disadvantages. They are:
\subsubsection{\label{sec:sec231}The economical model}
Introduced in the literature in Ref.~\cite{b3} and further analyzed in Refs.~\cite{b4,dong}. It makes use of only two scalar triplets, which together with their vacuum expectation values (VEV) are:
\begin{subequations}
\label{1aa}
\begin{align}
\label{1}
 \Phi_1(1,3^*,-1/3)&=
\begin{pmatrix}
 \phi_1^-\\
\phi_1^0\\
\phi_1^{\prime 0}
\end{pmatrix},\:\textrm{with VEV:}\:
\langle\Phi_1\rangle=
\frac{1}{\sqrt{2}}
\begin{pmatrix}
 0\\
v_1\\
V_1
\end{pmatrix},\\
\label{3}
\Phi_3(1,3^*,2/3)&=
\begin{pmatrix}
 \phi_3^0\\
\phi_3^+\\
\phi_3^{\prime+}
\end{pmatrix},\:\textrm{with VEV:}\:
\langle\Phi_3\rangle=
\frac{1}{\sqrt{2}}
\begin{pmatrix}
v_3\\
0\\
0
\end{pmatrix}.
\end{align}
\end{subequations}

The former structure is the simplest one, able to break the 3-3-1 symmetry down to $U(1)_{Q}$ in a consistent way~\cite{b3}. In spite of its simplicity, it has the disadvantage of being unable to produce a consistent Fermion mass spectrum at tree-level. The claim in Ref.~\cite{dong} is that the quantum fluctuations can generate non-zero mass terms for all the Fermion fields, but a systematic (tedious) numerical analysis reproducing the fermion mass spectrum has not been published yet, although probably, the most serious hurdle for the survival of this model, is the existence of flavor changing neutral currents (FCNC) at tree-level, mediated by the Higgs scalar fields (only two sets of scalar fields producing masses for three Fermion families), neutral currents that severely constraint the parameters of the model.

For this economical model, the study of the scalar potential, using the method introduced in Refs.~\cite{b1,b2} (and briefly review in appendix A) has been presented in full detail in Ref.~\cite{yith}.

\subsubsection{\label{sec:sec232}The set with three scalar triplets}
This set makes use of the two scalar Higgs fields of the economical model, plus the extra one 
\begin{equation}\label{2aa}
\Phi_2(1,3^*,-1/3)=
\begin{pmatrix}
 \phi_2^-\\
\phi_2^0\\
\phi_2^{\prime 0}
\end{pmatrix},\:\textrm{with VEV:}\:
\langle\Phi_2\rangle=
\frac{1}{\sqrt{2}}
\begin{pmatrix}
 0\\
v_2\\
V_2
\end{pmatrix},\\
\end{equation}
where we have assumed that all the five electrically neutral components do acquire non-zero VEV. This set of three scalar Higgs fields, with the vacuum aligned such that $V_1=v_2=0$, was used for the first time in Refs.~\cite{vl,ozer}. The set 
has the advantage of being able to produce tree-level masses for all the Fermion fields (which is true even for the particular alignment used in the original papers), but it can not completely avoid the presence of FCNC at tree-level, coming from the scalar sector.

Notice that the VEV $\langle\Phi_1\rangle$ and $\langle\Phi_2\rangle$ generate masses for the exotic quarks and the new heavy gauge bosons, while VEV  $\langle\Phi_3\rangle$ generates masses for ordinary fermions and for the SM gauge bosons. To keep the model consistent with low energy phenomenology, in this paper we will use $\langle\Phi_3\rangle\neq 0$ and the hierarchy
\begin{equation}\label{5}
 V_1,V_2\gg v_1,v_2,v_3,
\end{equation}
except for those cases when $V_1$ or $V_2$ are zero, when the hierarchy becomes
\begin{equation}\label{5p}
 V_i\gg v_1,v_2,v_3; \;\;\; i=1,2.
\end{equation}
(Taking $\langle\Phi_3\rangle=0$ implies that several fermion fields remain massless.)

\subsubsection{\label{sec:sec233}The extended scalar set}
Introduced in the literature  in Refs.~\cite{huner}, it consists of four scalar triplets: $\Phi_1, \; \Phi_2$ and $\Phi_3$ as above with the vacuum aligned such that $V_1=v_2=0$ as in the original papers, plus a new scalar Higgs field 
\begin{equation}
\Phi_4(1,3^*,-1/3)=
\begin{pmatrix}
 \phi_2^-\\
\phi_2^0\\
\phi_2^{\prime 0}
\end{pmatrix},\:\textrm{with VEV:}\:
\langle\Phi_2\rangle=
\frac{1}{\sqrt{2}}
\begin{pmatrix}
 0\\
0\\
v_4
\end{pmatrix},\\
\end{equation}
with the hierarchy $v_1\sim v_3\sim v_4<<V_2\sim 1$ TeV. This set of four scalar fields combined with a convenient discrete symmetry~\cite{huner}, is able to generate several see-saw mechanisms, the basis of a consistent Fermion mass spectrum, and avoid, at the same time, tree-level FCNC coming from the scalar sector.

In the following analysis we will concentrate only in the set with 3 scalar triplets as defined in Sects.~(\ref{sec:sec231}) and~(\ref{sec:sec232}), with the most general VEV structure, but with the constraint derived in Appendix B.
\section{\label{sec:sec3}The scalar Potential}
The most general scalar potential which is 3-3-1 invariant, for the set of three scalar triplets $\Phi_1,\;\Phi_2$ and $\Phi_3$ is given by 

{\footnotesize
\begin{equation}\label{7a}
\begin{split}
&V^\prime(\Phi_1,\Phi_2,\Phi_3)=\mu_1^2\Phi_1^\dag\Phi_1+\mu_2^2\Phi_2^\dag\Phi_2+\mu_3^2\Phi_3^\dag\Phi_3
\\&
+\frac{1}{2}(\mu_4^2\,\Phi_1^\dag\Phi_2+{\mu_4^2}^*\,\Phi_2^\dag\Phi_1)+\lambda_1(\Phi_1^\dag\Phi_1)^2+\lambda_2(\Phi_2^\dag\Phi_2)^2
\\&
+\lambda_3(\Phi_3^\dag\Phi_3)^2
+\lambda_4(\Phi_1^\dag\Phi_1)(\Phi_2^\dag\Phi_2)
+\lambda_5(\Phi_1^\dag\Phi_1)(\Phi_3^\dag\Phi_3)
\\&
+\lambda_6(\Phi_2^\dag\Phi_2)(\Phi_3^\dag\Phi_3)
+\lambda_7(\Phi_1^\dag\Phi_2)(\Phi_2^\dag\Phi_1)
\\ &
+\lambda_8(\Phi_1^\dag\Phi_3)(\Phi_3^\dag\Phi_1)
+\lambda_9(\Phi_2^\dag\Phi_3)(\Phi_3^\dag\Phi_2)
\\ &+
(f\,\Phi_1^\dag\Phi_2+f^{*}\,\Phi_2^\dag\Phi_1)^2 
+\frac{1}{2}(\lambda_{11}\,\Phi_1^\dag\Phi_2+\lambda_{11}^*\Phi_2^\dag\Phi_1)(\Phi_1^\dag\Phi_1)
\\ &
+\frac{1}{2}(\lambda_{12}\,\Phi_1^\dag\Phi_2
+\lambda_{12}^*\Phi_2^\dag\Phi_1)(\Phi_2^\dag\Phi_2) 
\\ &
+\frac{1}{2}(\lambda_{13}\,\Phi_1^\dag\Phi_2+\lambda_{13}^*\Phi_2^\dag\Phi_1)(\Phi_3^\dag\Phi_3)\\ 
&+\frac{1}{2}\left[\lambda_{14}\,(\Phi_1^\dag\Phi_3)(\Phi_3^\dag\Phi_2)+\lambda_{14}^*\,(\Phi_3^\dag\Phi_1)
(\Phi_2^\dag\Phi_3)\right] 
\\ &
+(g\,\epsilon_{ijk}\,\Phi_1^i\,\Phi_2^j\,\Phi_3^k+h.c.).
\end{split}
\end{equation}}
%
Since $\mu_4^2,\; f,\; \lambda_{11},\;\lambda_{12},\;\lambda_{13},\;\lambda_{14}$ and the trilinear coupling constant $g$ can be complex numbers, there are 26 free parameters in $V^\prime(\Phi_1,\Phi_2,\Phi_3)$ and 5 VEV, in principle all of them different from zero. The last element of $V^\prime$ correspond to the so called cubic term of the potential, which is closely related to a determinant function of Higgs fields due to the Levi-Civita component $\epsilon_{ijk}$.

For the sake of simplicity we are going to assume real VEV throughout this paper, which means that spontaneous CP violation is not going to be considered in our analysis. Notice also that the most general scalar potential $V^\prime(\Phi_1,\Phi_2,\Phi_3)$ in (\ref{7a}) is invariant under the local Gauge group $SU(3)_L\otimes U(1)_X$, invariance that is spontaneous broken by the VEV in $\langle\Phi_i\rangle,\; i=1,2,3$ down to $U(1)_Q$, where $Q$ is the electric charge generator in equation~(\ref{qele}). So, after the breaking of the symmetry, a consistent model will emerge only if eight massless Goldstone Bosons show up, coming from the transformed potential obtained from (\ref{7a}); zero mass Bosons that should be eaten up by the Gauge Bosons associated with the $SU(3)_L\otimes U(1)_X$ broken symmetry.

The scalar potential in Eq.~(\ref{7a}) is quite complicated and very difficult to study in a systematic way, and as far as we know, it has not been studied in full detail in the literature yet (and we do not intend to do it here either). A partial analysis of this general  potential, for the particular vacuum alignment $V_1=v_2=0$, has been done in Ref.~\cite{saro}. However, as mentioned in Refs.~\cite{folo,bb3_1}, by introducing discrete symmetries, the form of the potential simplifies largely and can be analyzed in detail, as we are going to do next.
\subsection{\label{sec:sec31}Discrete symmetry in the scalar potential}
Under assumption of the discrete symmetry $\Phi_1\rightarrow -\Phi_1$,
the most general potential obtained from (\ref{7a}), is presented in Appendix C, where it is demonstrated that $f$ can be taken as a single parameter. As a consequence of this, the reduced potential 
%
{\footnotesize
\begin{equation}
\label{10a}
\begin{split}
 V(\Phi_1,\Phi_2,\Phi_3)&=\mu_1^2\Phi_1^\dag\Phi_1+\mu_2^2\Phi_2^\dag\Phi_2+\mu_3^2\Phi_3^\dag\Phi_3
\\&
+\lambda_1(\Phi_1^\dag\Phi_1)^2
+\lambda_2(\Phi_2^\dag\Phi_2)^2+\lambda_3(\Phi_3^\dag\Phi_3)^2
\\&
+\lambda_4(\Phi_1^\dag\Phi_1)(\Phi_2^\dag\Phi_2)
+\lambda_5(\Phi_1^\dag\Phi_1)(\Phi_3^\dag\Phi_3)
\\&
+\lambda_6(\Phi_2^\dag\Phi_2)(\Phi_3^\dag\Phi_3)
+\lambda_7(\Phi_1^\dag\Phi_2)(\Phi_2^\dag\Phi_1)\\
&+\lambda_8(\Phi_1^\dag\Phi_3)(\Phi_3^\dag\Phi_1)
+\lambda_9(\Phi_2^\dag\Phi_3)(\Phi_3^\dag\Phi_2)
\\&+
\frac{\lambda_{10}}{2}(\Phi_1^\dag\Phi_2+\Phi_2^\dag\Phi_1)^2,
\end{split}
\end{equation}}
%
\hspace{-1mm}contains only 13 free parameters (instead of 26) and does not include the cubic term $\epsilon_{ijk}\Phi_1^i\Phi_2^j\Phi_3^k$. The rest of the paper will be dedicated to study this potential~(\ref{10a}).

A careful analysis shows now that, due to the absence of the cubic term, the potential $V(\Phi_1,\Phi_2,\Phi_3)$ turns out to be $U(3)\otimes U(1)_X$ invariant [instead of $SU(3)\otimes U(1)_X$] with the consequence that the most general VEV breaks this symmetry down to $U(1)_Q$ as before, producing now nine Goldstone Bosons, instead of the eight that can be Gauged away, due to the fact that the generator $I_3=Dg(1,1,1)$ gets also broken. This leaves an (unphysical?) extra zero mass scalar after the implementation of the Higgs mechanism. The simplest way to avoid this situation is by restoring the cubic term in the scalar potential (a dynamical breaking of the $U(3)$ symmetry), something does not allowed by the discrete symmetry imposed. But as shown in appendix (B), for the case when $\langle\Phi_3\rangle\neq 0$ the problem can be solved by demanding that $\langle\Phi_1\rangle$ and $\langle\Phi_2\rangle$ became linearly dependent (LD); this avoids an spontaneous breaking of the $U(3)\otimes U(1)_X$ symmetry down to $SU(3)\otimes U(1)_X$, with the consequence that the VEV must satisfy the constraint 
\begin{equation}\label{ldvev}
v_2V_1=v_1V_2,
\end{equation}
which can be used to express at least one VEV in terms of the rest. 

Notice that if $\langle\Phi_3\rangle=0$, the $U(3)$ generator $Diag.(1,0,0)$ remains unbroken by $\langle\Phi_1\rangle\oplus \langle\Phi_2\rangle\oplus\langle\Phi_3\rangle$, restoring in this way the eight Goldstone Bosons required. But we are not going to consider this unphysical situation as previously mentioned.

Before continuing, let us emphasize that constraint (\ref{ldvev}) is a consequence of demanding a consistent implementation of the Higgs mechanism for the breaking of the original $SU(3)\otimes U(1)_X$ local Gauge symmetry, respecting the electromagnetic $U(1)_Q$ invariance, and it is not coming from the minimization of the scalar potential. On the contrary, this constraint is taking into account when we study the stability and minimization of the potential.

Notice that first two papers in Ref.~\cite{vl} and  all papers in Ref.~\cite{folo}, the reduced potential $V(\Phi_1,\Phi_2,\Phi_3)$ was studied using the particular vacuum alignment $V_1=v_2=0$, with $V_2\gg v_1\neq 0$, in clear contradiction with equation (\ref{ldvev}). As an immediate consequence, in those papers an extra zero mass Goldstone Boson which cannot be Gauged away appears, making the analysis and some of the conclusions in all those papers dubious. To add in proof, notice that the four papers in Ref.~\cite{bb3_1} make use of that extra Goldstone Boson to implement the Peccei-Quinn symmetry~\cite{pq} in the context of the 3-3-1 model with right handed neutrinos, with the inconvenience of having in their analysis an unrealistic axion that is hidden by the introduction of an extra scalar field.

In what follows we are going to study the consistency of the scalar potential $V(\Phi_1,\Phi_2,\Phi_3)$ in Eq.~(\ref{10a}), under the linear dependent~(LD) constraint equation: $v_2V_1=v_1V_2$.

To start with, let us define as usual the scalar fields in the way:
\begin{subequations}\label{newfields}
\begin{align}
\phi_1^0&=\frac{v_1+H_1+iA_1}{\sqrt{2}},\;\; \phi_1^{\prime 0}=\frac{V_1+H^\prime_1+iA^\prime_1}{\sqrt{2}},\\
\phi_2^0&=\frac{v_2+H_2+iA_2}{\sqrt{2}},\;\; \phi_2^{\prime 0}=\frac{V_2+H^\prime_2+iA^\prime_2}{\sqrt{2}},\\
\phi_3^0&=\frac{v_3+H_3+iA_3}{\sqrt{2}},
\end{align}
\end{subequations}
where a real part $H$ is called in the literature a CP-even scalar and an imaginary part $A$ a CP-odd scalar or pseudoscalar field.
\subsection{\label{sec:sec32}Independent vacuum structures}
Assuming for the VEV the hierarchy in (\ref{5}) or in (\ref{5p}), and using the LD constraint relation (\ref{ldvev}), we classify in Table~\ref{c1}, all the possible 3-3-1 vacuum structures of $\Phi_1$ and $\Phi_2$, the two scalar triplets with identical quantum numbers, where at least one VEV is different from zero.

A careful analysis shows that not all the nine structures are independent. As a matter of fact, by performing an $SU(3)_L$ transformation on $\langle\Phi_1\rangle$ and on  $\langle\Phi_2\rangle$ in structure 1 of Table~\ref{c1}, we can obtain either the structure configuration 2 or the structure configuration 5. But it is not possible to make an $SU(3)_L\otimes U(1)_X$ transformation followed by a change of basis of the Higgs fields $\Phi_i\rightarrow\Phi_i^\prime$ of the form
\begin{equation}
\label{1c}
 \begin{pmatrix}
  \Phi_1^\prime\\
\Phi_2^\prime
 \end{pmatrix}=
U\begin{pmatrix}
  \Phi_1\\
\Phi_2
 \end{pmatrix},
\end{equation}
where $U$ is a $2\times2$ unitary matrix, such that the configuration 3 can be obtained; this is because the transformation~(\ref{1c}) violates the discrete symmetry $\Phi_1\rightarrow-\Phi_1$ previously imposed on the scalar potential. It is also possible to show that structures 3 and 4 are equivalent to each other due to the symmetry of the potential under the exchange $\Phi_1\leftrightarrow \Phi_2$, with some parameters renamed appropriately. In conclusion, the analysis shows that only structures 1 and 3 in Table~\ref{c1} are independent and are the only cases we are going to consider in our analysis.
\begin{table*}[ht]
\caption{Different VEV structures for scalars $\Phi_1$ and $\Phi_2$.}\label{c1}
\centering
\begin{tabular}{|c|c|c|}\hline
Structure & VEV & Vacuum alignments\\ \hline \hline&\\[-4.3mm]
& & \\
1& \quad $\begin{matrix}
          v_1,V_1,v_2,V_2\ne0\\
\textrm{and}\quad v_2\,V_1=v_1\,V_2
         \end{matrix}
$
  & $\langle\Phi_1\rangle\propto\begin{pmatrix}
                                                 0&v_1 &V_1
                                                \end{pmatrix}^T,\:
\langle\Phi_2\rangle\propto\begin{pmatrix}
                                                 0&v_2 &V_2
                                                \end{pmatrix}^T$\\ \hline&\\[-4.3mm]
& & \\
2 & \quad  $v_1,v_2=0;V_1,V_2\ne0$ & $\langle\Phi_1\rangle\propto\begin{pmatrix}
                                                 0&0 &V_1
                                                \end{pmatrix}^T,\:
\langle\Phi_2\rangle\propto\begin{pmatrix}
                                                0& 0 &V_2
                                                \end{pmatrix}^T$\\ \hline&\\[-4.3mm]
& & \\
3 &  \quad$v_2,V_2=0;v_1,V_1\ne0$ & $\langle\Phi_1\rangle\propto\begin{pmatrix}
                                                0& v_1 &V_1
                                                \end{pmatrix}^T,\:
\langle\Phi_2\rangle=\begin{pmatrix}
                                                0& 0 &0
                                                \end{pmatrix}^T$\\\hline&\\[-4.3mm]
& & \\
4 & \quad $v_1,V_1=0;v_2,V_2\ne0$ & $\langle\Phi_1\rangle=\begin{pmatrix}
                                                 0&0 &0
                                                \end{pmatrix}^T,\:
\langle\Phi_2\rangle\propto\begin{pmatrix}
                                                0& v_2 &V_2
                                                \end{pmatrix}^T$\\\hline&\\[-4.3mm]
& & \\
5 &  \quad$V_1,V_2=0;v_1,v_2\ne0$ & $\langle\Phi_1\rangle\propto\begin{pmatrix}
                                                 0&v_1 &0
                                                \end{pmatrix}^T,\:
\langle\Phi_2\rangle\propto\begin{pmatrix}
                                                 0&v_2 &0
                                                \end{pmatrix}^T$\\\hline&\\[-4.3mm]
& & \\
6 & \quad$v_2,V_2,V_1=0;v_1\ne0$ & $\langle\Phi_1\rangle\propto\begin{pmatrix}
                                                0& v_1 &0
                                                \end{pmatrix}^T,\:
\langle\Phi_2\rangle=\begin{pmatrix}
                                                0& 0 &0
                                                \end{pmatrix}^T$\\\hline&\\[-4.3mm]
& & \\
7 &  \quad$v_2,V_2,v_1=0;V_1\ne0$ & $\langle\Phi_1\rangle\propto\begin{pmatrix}
                                                0& 0 &V_1
                                                \end{pmatrix}^T,\:
\langle\Phi_2\rangle=\begin{pmatrix}
                                                0& 0 &0
                                                \end{pmatrix}^T$\\\hline&\\[-4.3mm]
& & \\
8 & \quad $v_1,V_1,V_2=0;v_2\ne0$ & $\langle\Phi_1\rangle=\begin{pmatrix}
                                                 0&0 &0
                                                \end{pmatrix}^T,\:
\langle\Phi_2\rangle\propto\begin{pmatrix}
                                                0& v_2 &0
                                                \end{pmatrix}^T$\\\hline&\\[-4.3mm]
& & \\
9 & \quad $v_1,V_1,v_2=0;V_2\ne0$ & $\langle\Phi_1\rangle=\begin{pmatrix}
                                                 0&0 &0
                                                \end{pmatrix}^T,\:
\langle\Phi_2\rangle\propto\begin{pmatrix}
                                                0& 0 &V_2
                                                \end{pmatrix}^T$\\\hline
 \end{tabular}
\end{table*}
\subsubsection{\label{sec:sec321}Vacuum structure with $v_1, V_1,v_2,V_2\neq0$}
It will be shown further below that, by minimization methods applied on potential, the scalars acquiring non-zero VEVs along their electrically neutral entries, is highly suggested~(\ref{123}).
{\scriptsize
\begin{equation}
\label{53}
\langle\Phi_1\rangle=
\frac{1}{\sqrt{2}}
\begin{pmatrix}
 0\\
v_1\\
V_1
\end{pmatrix},\:
\langle\Phi_2\rangle=
\frac{1}{\sqrt{2}}
\begin{pmatrix}
 0\\
v_2\\
V_2
\end{pmatrix},\:
\langle\Phi_3\rangle=
\frac{1}{\sqrt{2}}
\begin{pmatrix}
v_3\\
0\\
0
\end{pmatrix},
\end{equation}}
\hspace{-1.5mm}with $v_2\,V_1=v_1\,V_2$.
And requiring that in the shifted potential obtained from $V(\Phi_1,\Phi_2,\Phi_3)$, the linear terms in the fields must be absent, we get in the tree-level approximation the following constraint equations
%
{\scriptsize
\begin{subequations}
\label{16a}
\begin{align}\label{9}
{2\mu_1}^2&+{\left( \lambda_7+\lambda_4+2\,\lambda_{10}\right) \,\left( {V_2}^{2}+{v_2}^{2}\right)
+2\,\lambda_1\,\left( {V_1}^{2}+{v_1}^{2}\right) +\lambda_5\,{v_3}^{2}}=0,\\ \label{10}
{2\mu_2}^2&+{2\,\lambda_2\,\left( {V_2}^{2}+{v_2}^{2}\right) +\left( \lambda_7+\lambda_4+2\,\lambda_{10}\right) \,\left( {V_1}^{2}+{v_1}^{2}\right) +\lambda_6\,{v_3}^{2}}=0,\\ \label{8}
{2\mu_3}^2&+{{\lambda_6}\,({V_2}^2+{ v_2}^2)+{\lambda_5}\,({V_1}^2+\,{ v_1}^2)+2\,{ \lambda_3}\,v_3^2}=0.
\end{align}
\end{subequations}}
\hspace{-1.5mm}where the VEVs must satisfy the constraint (\ref{ldvev}). In section~\ref{sec:sec53}, we will express $\{v_1^2+V_1^2,v_2^2+V_2^2,v_3^2\}$ in terms of the parameters of the potential by using the orbital variables method.
\paragraph{Spectrum in the scalar neutral sector}
\label{sec:sec3211}
In the $H_1,H_2,H_1^\prime$, $H_2^\prime,H_3$ basis, the square mass matrix can be calculated by using $M_{ij}^2=\left.[\partial  V(\Phi_1,\Phi_2,\Phi_3)/\partial H_i\partial H_j]\right|_{fields=0}$. After imposing constraints~(\ref{16a}), we get $M^2_H=$
%
{\scriptsize
\begin{equation*}
\label{11}
\begin{split}
&\left(
\begin{matrix}
2(\,u_5\,{V_2}^{2}+\,\lambda_{1}\,{v_1}^{2}) & \left( \lambda_{4}-4\,u_5\right) \,v_1\,v_2-2\,u_5\,V_1\,V_2 
\cr 
\left( \lambda_{4}-4\,u_5\right) \,v_1\,v_2-2\,u_5\,V_1\,V_2 & 2(\,u_5\,{V_1}^{2}+\,\lambda_{2}\,{v_2}^{2}) 
\cr 
2(\,\lambda_{1}\,v_1\,V_1-\,u_5\,v_2\,V_2) & \left( \lambda_{4}-2\,u_5\right) \,v_2\,V_1 
\cr
\left( \lambda_{4}-2\,u_5\right) \,v_2\,V_1 & 2(\,\lambda_{2}\,v_2\,V_2-\,u_5\,v_1\,V_1) 
\cr 
\lambda_{5}\,v_3\,v_1 & \lambda_{6}\,v_3\,v_2
\end{matrix}\right.
\\
&\left.
\begin{matrix}
2(\,\lambda_{1}\,v_1\,V_1-\,u_5\,v_2\,V_2) & \left( \lambda_{4}-2\,u_5\right) \,v_2\,V_1 & \lambda_{5}\,v_3\,v_1
\cr 
\left( \lambda_{4}-2\,u_5\right) \,v_2\,V_1 & 2(\,\lambda_{2}\,v_2\,V_2-\,u_5\,v_1\,V_1) & \lambda_{6}\,v_3\,v_2
\cr 
 2(\,\lambda_{1}\,{V_1}^{2}+\,u_5\,{v_2}^{2}) & \left( \lambda_{4}-4\,u_5\right) \,V_1\,V_2-2\,u_5\,v_1\,v_2 & \lambda_{5}\,v_3\,V_1
\cr
 \left( \lambda_{4}-4\,u_5\right) \,V_1\,V_2-2\,u_5\,v_1\,v_2 & 2(\,\lambda_{2}\,{V_2}^{2}+\,u_5\,{v_1}^{2}) & \lambda_{6}\,u\,V_2
\cr 
\lambda_{5}\,v_3\,V_1 & \lambda_{6}\,u\,V_2 & 2\,\lambda_{3}\,{v_3}^{2}
\end{matrix}\right)
\end{split}
\end{equation*}}
\hspace{-1.3mm}where the term $u_5=-(\lambda_7+2\lambda_{10})/4$ has been used. This mass matrix has zero determinant providing us with a Goldstone Boson $G_1$ and four massive scalar fields. The analytic mass values are not easy to find, but in the approximation $V_1\sim V_2\gg v_1\sim v_2\sim v_3$ they are
{\scriptsize
\begin{align}
\begin{split}
M_{he_1}^2&\approx
{ \lambda_2}\,
 { V_2}^2+{ \lambda_1}\,{ V_1}^2\\
&+\sqrt{\left({ \lambda_1}\,{ V_1}^2-{ \lambda_2}\,{ V_2}^2
 \right)^2+\left(\lambda_4+\lambda_7+2\,\lambda_{10}\right)^2
\,{ V_1}^2\,{ V_2}^2},
\end{split}
\\
\begin{split}
M_{he_2}^2&\approx
{ \lambda_2}\,
 { V_2}^2+{ \lambda_1}\,{ V_1}^2\\
&-\sqrt{\left({ \lambda_1}\,{ V_1}^2-{ \lambda_2}\,{ V_2}^2
 \right)^2+\left(\lambda_4+\lambda_7+2\,\lambda_{10}\right)^2\,{ V_1}^2\,{ V_2}^2},
\end{split}
\\ \label{cite}
M_{he_3}^2&\approx
-{\frac{\left({ \lambda_7}+2\,{ \lambda_{10}}\right)\,\left({ V_2}^2+ { V_1}^2\right)}{2}},\\
M_{he_4}^2&\approx 2\lambda_3\,v_3^2\quad\textrm{with}\quad he_4\approx H_3,
\end{align}}
\hspace{-1.8mm}where the scalar $he_4$  is light and can be identified as the SM Higgs Boson scalar. In order to have positive masses for all the former scalars, the following constraints must be satisfied

\begin{subequations}
\label{20}
\begin{align}
\label{mases}
\lambda_1,\,\lambda_2,\lambda_3&>0,\\ 
\label{masesp}
4\lambda_1\,\lambda_2&>(\lambda_4+\lambda_7+2\,\lambda_{10})^2,\\ 
\label{masespp}
\left({ \lambda_7}+2\,{ \lambda_{10}}\right)&<0.
\end{align}
\end{subequations}
\paragraph{\label{sec:sec3212} Spectrum in the pseudoscalar neutral sector}
In the $A_1,A_2,A_1^\prime,A_2^\prime,A_3$ basis the square mass matrix is given by $M^2_A=$
%
{\scriptsize
\begin{equation*}
\label{12}
\begin{split}
&\left(\begin{matrix}
2\,u_5\,{V_2}^{2}-\lambda_{10}\,{v_2}^{2} & \lambda_{10}\,v_1\,v_2-2\,u_5\,V_1\,V_2 & \frac{\lambda_{7}\,v_2\,V_2}{2}  
\cr 
\lambda_{10}\,v_1\,v_2-2\,u_5\,V_1\,V_2 & 2\,u_5\,{V_1}^{2}-\lambda_{10}\,{v_1}^{2} & -\frac{\lambda_{7}\,v_2\,V_1}{2}  
\cr 
\frac{\lambda_{7}\,v_2\,V_2}{2} & -\frac{\lambda_{7}\,v_2\,V_1}{2} & 2\,u_5\,{v_2}^{2}-\lambda_{10}\,{V_2}^{2}   
\cr 
-\frac{\lambda_{7}\,v_2\,V_1}{2} & \frac{\lambda_{7}\,v_1\,V_1}{2} & \lambda_{10}\,V_1\,V_2-2\,u_5\,v_1\,v_2 
\cr 0 & 0 & 0  
\end{matrix}\right.
\\
&\left.\begin{matrix}
 -\frac{\lambda_{7}\,v_2\,V_1}{2} & 0\\
 \frac{\lambda_{7}\,v_1\,V_1}{2} & 0\\
\lambda_{10}\,V_1\,V_2-2\,u_5\,v_1\,v_2 & 0\\
 2\,u_5\,{v_1}^{2}-\lambda_{10}\,{V_1}^{2} & 0\\
0 & 0
\end{matrix}\right)
\end{split}
\end{equation*}}
\hspace{-1.6mm}which is a rank-2 matrix, giving three Goldstone Bosons and two heavy pseudoscalar particles with masses given by 
\begin{align}
 M_{ho_1}^2&\approx -\frac{(\lambda_7+2\lambda_{10})(V_1^2+V_2^2)}{2},\\ \label{mco}
M_{ho_2}^2&\approx -\lambda_{10}(V_1^2+V_2^2),
\end{align}
where $M_{ho_1}^2>0$ due to the constraint~(\ref{masespp})], and the condition  $M_{ho_2}^2>0$ implies the new constraint 
\begin{equation}
\label{24}
\lambda_{10}<0.
\end{equation}
%
\paragraph{\label{sec:sec3213} Spectrum in the charged scalar sector}
In the $\phi_1^\pm,\phi_2^\pm,\phi_3^\pm$, $\phi_3^{\prime\pm}$ basis, the square mass matrix is given by $M^2_\phi=$
{\footnotesize
\begin{equation}
\label{13}
\begin{split}
\frac{1}{2}
&\left(\begin{matrix}
4\,u_5\,\left( {V_2}^{2}+{v_2}^{2}\right) +\lambda_{8}\,{v_3}^{2} & -4\,u_5\,\left( V_1\,V_2+v_1\,v_2\right)  
\cr 
-4\,u_5\,\left( V_1\,V_2+v_1\,v_2\right)  & 4\,u_5\,\left( {V_1}^{2}+{v_1}^{2}\right) +\lambda_{9}\,{v_3}^{2} 
\cr 
\lambda_{8}\,v_3\,v_1 & \lambda_{9}\,v_3\,v_2 
\cr 
\lambda_{8}\,v_3\,V_1 & \lambda_{9}\,v_3\,V_2 
\end{matrix}\right.
\\
&\left.\begin{matrix}
        \lambda_{8}\,v_3\,v_1 & \lambda_{8}\,v_3\,V_1
\cr 
 \lambda_{9}\,v_3\,v_2 & \lambda_{9}\,v_3\,V_2
\cr 
\lambda_{9}\,{v_2}^{2}+\lambda_{8}\,{v_1}^{2} & \lambda_{9}\,v_2\,V_2+\lambda_{8}\,v_1\,V_1
\cr 
\lambda_{9}\,v_2\,V_2+\lambda_{8}\,v_1\,V_1 & \lambda_{9}\,{V_2}^{2}+\lambda_{8}\,{V_1}^{2}
      \end{matrix}\right)
\end{split}
\end{equation}}
%
\hspace{-.9mm}with $u_5$ defined above. The mass matrix~(\ref{13}) is a rank-2 matrix, implying the existence of four Goldstone Bosons and four massive charged scalars, with masses given by 
\begin{align}
\label{mh1pm}
M_{h^\pm_1}^2&\approx \frac{\lambda_8V_1^2+\lambda_9V_2^2}{2}>0,\\ 
\label{mh2pm}
M_{h_2^\pm}^2&\approx -\frac{(\lambda_7+2\lambda_{10})(V_1^2+V_2^2)}{2},
\end{align}
where again we have $M_{h_2^\pm}^2>0$ due to the constraint (\ref{masespp}).

Counting Goldstone Bosons we have a total of eight: an scalar and three pseudoscalars which are used to provide with masses to four electrically neutral gauge bosons ($Z^0,\;Z^{\prime 0},\;K^0$ and $\bar{K}^0$), and four charged ones which are used to provide with masses to $W^\pm$ and to $K^\pm$. This shows the consistency of our analysis.

\subsubsection{\label{sec:sec322}Vacuum structure with $v_2=V_2=0,v_1,V_1\neq0$}
In this section we are going to study the other independent structure given in Table~\ref{c1}, where the LD between $\langle\Phi_1\rangle$ and $\langle\Phi_2\rangle$ must be respected. The VEV configuration structure  is
{\scriptsize
\begin{equation}
\label{53a}
\langle\Phi_1\rangle=
\frac{1}{\sqrt{2}}
\begin{pmatrix}
 0\\
v_1\\
V_1
\end{pmatrix},\:
\langle\Phi_2\rangle=
\frac{1}{\sqrt{2}}
\begin{pmatrix}
 0\\
0\\
0
\end{pmatrix},\:
\langle\Phi_3\rangle=
\frac{1}{\sqrt{2}}
\begin{pmatrix}
v_3\\
0\\
0
\end{pmatrix}.
\end{equation}}
\hspace{-.9mm}In the tree level approximation, the constraint equations are now 
\begin{subequations}
\label{54}
\begin{align}
\mu_1^2&+\lambda_1\,(V_1^2+v_1^2)+\frac{\lambda_5}{2}\,v_3^2=0,\\
\mu_3^2&+\frac{\lambda_5}{2}\,(V_1^2+v_1^2)+\lambda_3\,v_3^2=0,
\end{align}
\end{subequations}
where there is no a constraint relation for $\mu_2^2$, which now becomes  a free parameter of the model.
\paragraph{\label{sec:sec3221}Spectrum in the scalar neutral sector}
The first result obtained is that the fields $H_1,H_1^\prime$ and $H_3$  do not mix with $H_2$ and $H_2^\prime$ 

In the $H_1,H_1^\prime,H_3$ basis, the square mass matrix is
\begin{equation}
M_{e2}^2=
 \begin{pmatrix}
2\,\lambda_1\,v_1^2 & 2\,\lambda_1\,v_1\,V_1 & \lambda_5\,v_1\,v_3\\
2\,\lambda_1\,v_1\,V_1 & 2\,\lambda_1\,V_1^2 & \lambda_5\,v_3\,V_1 \\
\lambda_5\,v_1\,v_3 & \lambda_5\,v_3\,V_1 & 2\,\lambda_3\,v_3^2
 \end{pmatrix},
\end{equation}
which is a rank-2 matrix, implying the existence of one Goldstone Boson.

Now, in the $H_2,H_2^\prime$ basis, the rank-2  mass matrix is.
{\footnotesize
\begin{equation}
M_{e3}^2=
\begin{pmatrix}
\frac{\lambda_4V_1^2+S\,v_1^2+\lambda_6v_3^2}{2}+\mu_2^2& \frac{(\lambda_7+2\lambda_{10})v_1V_1}{2} \\
\frac{(\lambda_7+2\lambda_{10})v_1V_1}{2} & \frac{S\,V_1^2+\lambda_4v_1^2+\lambda_6v_3^2}{2}+\mu_2^2
 \end{pmatrix},
\end{equation}}
\hspace{-1.4mm}where $S=\lambda_4+\lambda_7+2\lambda_{10}$.
For this vacuum structure the analytic scalar square mass values can be calculated exactly; they are 
%
{\footnotesize
\begin{align}
\label{47}
\begin{split}
M_{he'_1}^2&=
\lambda_1\,(v_1^2+V_1^2)+\lambda_3\,v_3^2\\
&+\sqrt{\left[\lambda_1(V_1^2+v_1^2)-\lambda_3v_3^2\right]^2+
\lambda_5^2v_3^2(V_1^2+v_1^2)},
\end{split}
\\ 
\label{48}
\begin{split}
M_{he'_2}^2&=
\lambda_1\,(v_1^2+V_1^2)+\lambda_3\,v_3^2
\\&-\sqrt{\left[\lambda_1(V_1^2+v_1^2)-\lambda_3v_3^2\right]^2+
\lambda_5^2v_3^2(V_1^2+v_1^2)},
\end{split}
\\ 
\label{50}
M_{he'_3}^2&=
\frac{(\lambda_7+\lambda_4+2\,\lambda_{10})\,(V_1^2+v_1^2)+\lambda_6\,v_3^2}{2}+\,\mu_2^2>0,
\\ \label{51}
M_{he'_4}^2&=
\frac{\lambda_4\,(V_1^2+v_1^2)+\lambda_6\,v_3^2}{2}+\,\mu_2^2>0.
\end{align}}
%
%
In order to have positive masses for the first two scalars, the following constraint equations must be satisfied:
\begin{equation}
\label{52}
 \lambda_1,\lambda_3>0\quad \textrm{and}\quad 4\,\lambda_1\,\lambda_3>\lambda_5^2.
\end{equation}
Notice by the way that the masses for the scalar fields~(\ref{47}) and (\ref{48}) correspond to the masses of the CP even physical fields in the economical model~\cite{b3,yith,dong}, and thus $he'_2$ can be identified as the SM Higgs boson scalar.
\paragraph{\label{sec:sec3222}Spectrum in the pseudoscalar neutral sector.}
In this sector the fields $A_1,A^\prime_1,A_3$ do not get mass entries, becoming automatically 3 odd Goldstone Bosons. Now, in the basis $A_2,A_2^\prime$ the rank-2 square mass matrix $M_0^2$ is:
{\small
\begin{equation*}
\begin{pmatrix}
\frac{\lambda_4\,V_1^2+(\lambda_7+\lambda_4)\,v_1^2+\lambda_6\,v_3^2}{2}+\mu_2^2 & \frac{\lambda_7\,v_1\,V_1}{2}\\
\frac{\lambda_7\,v_1\,V_1}{2} & \frac{(\lambda_7+\lambda_4)\,V_1^2+\lambda_4\,v_1^2+\lambda_6\,v_3^2}{2}+\mu_2^2
 \end{pmatrix},
\end{equation*}}
with eigenvalues for the physical fields given by
\begin{align}
\label{54b}
M_{ho'_1}^2&=\frac{(\lambda_7+\lambda_4)(V_1^2+v_1^2)+\lambda_6v_3^2}{2}+\mu_2^2>0,\\
\label{55b}
M_{ho'_2}^2&= \frac{\lambda_4(V_1^2+v_1^2)+\lambda_6v_3^2}{2}+\mu_2^2>0.
\end{align}
\paragraph{\label{sec:sec3223}Spectrum in the charged scalar sector}
In the $\phi_1^\pm, \phi_2^\pm, \phi_3^\pm$, $\phi_3^{\prime\pm}$ basis the $4\times 4$ square mass matrix $M_c^2$ is
{\small
\begin{equation*}
\begin{pmatrix}
\frac{\lambda_8\,v_3^2}{2} &0& \frac{(\lambda_8\,v_1\,v_3)}{2}& \frac{\lambda_8\,v_3\,V_1}{2}\\
0& \frac{\lambda_4\,V_1^2+\lambda_4\,v_1^2+(\lambda_9+\lambda_6)\,v_3^2+2\,\mu_2^2}{2} & 0&0\\
\frac{\lambda_8\,v_1\,v_3}{2} &0&\frac{\lambda_8\,v_1^2}{2}&\frac{ \lambda_8\,v_1\,V_1}{2}\\
\frac{\lambda_8\,v_3\,V_1}{2} & 0& \frac{\lambda_8\,v_1\,V_1}{2} & \frac{\lambda_8\,V_1^2}{2}
 \end{pmatrix},
\end{equation*}}
\hspace{-1.6mm}which is a rank-2 mass matrix producing in this way four Goldstone Bosons. The remaining physical fields have square masses:
\begin{align}
\label{55}
M_{h_1^{\prime\pm}}^2&=\frac{\lambda_8(V_1^2+v_1^2+v_3^2)}{2},\\
\label{58a}
M_{h_2^{\prime\pm}}^2&=\frac{\lambda_4(V_1^2+v_1^2)+(\lambda_9+\lambda_6)v_3^2}{2}+\mu_2^2>0,
\end{align}
where $h_2^{\prime\pm}=\phi_2^\pm$. Now, for $M_{h_1^{\prime\pm}}^2>0$ it must hold
\begin{equation}
\label{59}
 \lambda_8>0.
\end{equation}
Notice again that the masses of the two physical charged scalars coincide with that masses in the economical 3-3-1 model.

Counting Goldstone Bosons we get again a consistent spectrum.

In the following two sections we are going to derive bounds on the parameters of the scalar potential~(\ref{10a}) that result from the following conditions:
\begin{itemize}
\item The potential $V(\Phi_1,\Phi_2,\Phi_3)$ must be stable,
\item The potential must be able to break the symmetry $SU(3)_L\otimes U(1)_X$ down to $U(1)_Q$, in a consistent way.
\end{itemize}
\section{\label{sec:sec4} Stability of the scalar Potential}
The scalar potential $V(\Phi_1,\Phi_2,\Phi_3)$ in (\ref{10a}) is stable if it is bounded from below; this guarantees the existence of a global minimum in the potential. The stability of the scalar potential turns out to be independent of the values taken by the VEV: $v_1,v_2,v_3,V_1$ and $V_2$, as it is going to be shown in the following analysis. In other words, the results obtained below are valid,  independent of the vacuum structure chosen.

\subsection{\label{sec:sec41}The orbital variables}
The most general gauge invariant and renormalizable scalar potential $V(\Phi_1,\Phi_2,\Phi_3)$ in (\ref{10a}), that does not contain the cubic term, for the three Higgs scalar triplets $\Phi_1,\Phi_2,$ and $\Phi_3$, is an  Hermitian linear combination of terms of the form
\begin{equation}
 \Phi_i^\dag\Phi_j,\quad (\Phi_i^\dag\Phi_j)(\Phi_k^\dag\Phi_l),
\end{equation}
where $i,j,k,l\in{1,2,3}$. 

Following the method presented in appendix A, it is convenient to discuss the properties of the scalar  potential, such as its stability and its spontaneous symmetry breaking, in terms of gauge invariant expressions. For this purpose we arrange the $SU(3)_L$ invariant scalar products into the the following three $2\times2$ hermitian matrices
%
\begin{equation}
\label{45}
\begin{split}
\underline K&=
\begin{pmatrix}
\Phi_1^\dag\Phi_1& \Phi_2^\dag\Phi_1\\
\Phi_1^\dag\Phi_2& \Phi_2^\dag\Phi_2
\end{pmatrix},
%
\quad\underline L=
\begin{pmatrix}
\Phi_1^\dag\Phi_1& \Phi_3^\dag\Phi_1\\
\Phi_1^\dag\Phi_3& \Phi_3^\dag\Phi_3
\end{pmatrix},
\\
\underline M&=
\begin{pmatrix}
\Phi_2^\dag\Phi_2& \Phi_3^\dag\Phi_2\\
\Phi_2^\dag\Phi_3& \Phi_3^\dag\Phi_3
\end{pmatrix},
\end{split}
\end{equation}
%
where according with Eq.~(\ref{6ab}), each matrix is related to the following four real parameters
{\scriptsize
\begin{subequations}
\label{71a}
\begin{align}
\label{71}
\underline K:\;&\left\{
\begin{matrix}
\Phi_1^\dag\Phi_1=(K_0+K_3)/2, &\Phi_2^\dag\Phi_2=(K_0-K_3)/2,\\
\Phi_1^\dag\Phi_2=(K_1+i\;K_2)/2, &\Phi_2^\dag\Phi_1=(K_1-i\;K_2)/2,
\end{matrix}\right.
\\
\label{72}
\underline L:\;&\left\{
\begin{matrix}
\Phi_1^\dag\Phi_1=(L_0+L_3)/2, &\Phi_3^\dag\Phi_3=(L_0-L_3)/2,\\
\Phi_1^\dag\Phi_3=(L_1+i\;L_2)/2, &\Phi_3^\dag\Phi_1=(L_1-i\;L_2)/2,
\end{matrix}\right.
\\
\label{73}
\underline M:\;&\left\{
\begin{matrix}
\Phi_2^\dag\Phi_2=(M_0+M_3)/2, &\Phi_3^\dag\Phi_3=(M_0-M_3)/2,\\
\Phi_2^\dag\Phi_3=(M_1+i\;M_2)/2, &\Phi_3^\dag\Phi_2=(M_1-i\;M_2)/2,
\end{matrix}\right.
\end{align}
\end{subequations}}
with the constraints
{\small
\begin{subequations}
\label{63}
\begin{align}
 K_0\ge0,&\quad K_0^2-K_1^2-K_2^2-K_3^2=K_0^2-\boldsymbol K^2\ge0,\\
\label{63b}
L_0\ge0,&\quad L_0^2-L_1^2-L_2^2-L_3^2=L_0^2-\boldsymbol L^2\ge0,\\
\label{63c}
M_0\ge0,&\quad M_0^2-M_1^2-M_2^2-M_3^2=M_0^2-\boldsymbol M^2\ge0.
\end{align}
\end{subequations}}
\hspace{-1.3mm}The scalar products $\Phi_1^\dag\Phi_1$, $\Phi_2^\dag\Phi_2$ and $\Phi_3^\dag\Phi_3$, present in the expressions~(\ref{71}), (\ref{72}) and~(\ref{73}), allow us to eliminate three of the 12 variables due to the fact that 
\begin{subequations}
\label{64}
\begin{align}
 K_3&=L_0-M_0,\\
\label{64b}
L_3&=K_0-M_0,\\
\label{64c}
M_3&=K_0-L_0,
\end{align}
\end{subequations}
ending up with only the following nine real orbital variables, used to describe the full scalar potential
\begin{equation}
\label{65}
K_0,L_0,M_0,K_1,K_2,
L_1,L_2,
M_1,M_2.
\end{equation}
With the help of the former variables, the scalar potential (\ref{10a}) may be written as 
{\footnotesize
\begin{equation}
\label{81}
V(\Phi_1,\Phi_2,\Phi_3)=(V_{2K}+V_{4K})+(V_{2L}+V_{4L})+(V_{2M}+V_{4M}),
\end{equation}}
\hspace{-1.3mm}where as can be seen, the general space splits as the direct sum of three subspaces, due to the particular simple form of the scalar potential in (\ref{10a}) and to the fact that $\langle\Phi_3\rangle$ is orthogonal to $\langle\Phi_1\rangle$  and to $\langle\Phi_2\rangle$, something which guarantees the validity of the generalized Schwartz's Inequality
\[\langle\Phi_1|\Phi_1\rangle\langle\Phi_2|\Phi_2\rangle\langle\Phi_3|\Phi_3\rangle\geq\langle\Phi_1|\Phi_2\rangle\langle\Phi_1|\Phi_3\rangle\langle\Phi_2|\Phi_3\rangle.\]
With the use of the real parameters $\xi_{k(l,m)0}$, $\xi_{k(l,m)a}$, \linebreak $\eta_{k(l,m)00}$, $\eta_{k(l,m)a}$ and  $\eta_{k(l,m)ab}=\eta_{k(l,m)ba}$, the following functions defined in 
the domain $|\boldsymbol k|,|\boldsymbol l|,|\boldsymbol m|\le1$.
\begin{subequations}\label{62a}
\begin{align}
J_{k2}(\boldsymbol k)&=\xi_{k0}+\boldsymbol \xi_{k}\cdot\boldsymbol k,\\
J_{k4}(\boldsymbol k)&=\eta_{k00}+2\boldsymbol \eta_{k}\cdot\boldsymbol k+\boldsymbol{k}\cdot E_k\cdot \boldsymbol{k},\\ \label{63a}
J_{l2}(\boldsymbol l)&=\xi_{l0}+\boldsymbol \xi_{l}\cdot\boldsymbol l,\\
J_{l4}(\boldsymbol l)&=\eta_{l00}+2\boldsymbol \eta_{l}\cdot\boldsymbol l+\boldsymbol{l}\cdot E_l\cdot \boldsymbol{l}, \\
J_{m2}(\boldsymbol m)&=\xi_{m0}+\boldsymbol \xi_{m}\cdot\boldsymbol m,\\ 
J_{m4}(\boldsymbol m)&=\eta_{m00}+2\boldsymbol \eta_{m}\cdot\boldsymbol m+\boldsymbol{m}\cdot E_m\cdot \boldsymbol{m},
\end{align}
\end{subequations}
where in according to~(\ref{63})
\begin{subequations}
\label{65aa}
\begin{align}
\boldsymbol k&=\boldsymbol K/K_0,\:\left(|\boldsymbol k|\le1\right);\\ 
\boldsymbol l&=\boldsymbol L/L_0,\:\left(|\boldsymbol l|\le1\right);\\ 
\boldsymbol m&=\boldsymbol M/M_0,\:\left(|\boldsymbol m|\le1\right),
\end{align}
\end{subequations}
for $K_0,L_0,M_0>0$, allows us to write the terms of potential~(\ref{81}).
\begin{subequations}
\label{79}
\begin{align}
V_{2K}&=\xi_{k0}K_0+\xi_{ka}K_a=K_0J_{k2}(\boldsymbol k),
\\
\begin{split}
V_{4K}&=\eta_{k00}K_0^2+2K_0\eta_{ka}K_a+K_a\eta_{kab}K_b
\\&=K_0^2J_{k4}(\boldsymbol k),
\end{split}
\\
V_{2L}&=\xi_{l0}L_0+\xi_{la}L_a=L_0J_{l2}(\boldsymbol l),
\\
\begin{split}
V_{4L}&=\eta_{l00}L_0^2+2L_0\eta_{la}L_a+L_a\eta_{lab}L_b
\\
&=L_0^2J_{l4}(\boldsymbol l),
\end{split}
\\
V_{2M}&=\xi_{m0}M_0+\xi_{ma}M_a=M_0J_{m2}(\boldsymbol m),
\\
\begin{split}
V_{4M}&=\eta_{m00}M_0^2+2M_0\eta_{ma}M_a+M_a\eta_{mab}M_b\\
&=M_0^2J_{m4}(\boldsymbol m),
\end{split}
\end{align}
\end{subequations}
where sum over the indices $a$ and $b$ from 1 to 3 must be understood.
In the former expressions, the following notation has been used:
$E_k=\eta_{kab}, \; E_l=\eta_{lab}, \;E_m=\eta_{mab}$.
The parametrization employed in Eqs.~(\ref{45})-(\ref{65}) should not invalidate the stability conditions (in the strong sense) as far as sufficient conditions are concerned (necessary and sufficient conditions should be affected). 

On the other hand,  the parameters given in~(\ref{65}) does not imply that the matrix arrangements~(\ref{45}) can be established. In that way, the parameters~(\ref{65}) may help in the procedure to find the stationary points in the scalar potential, but it is necessary to verify at the end, if all  matrices~(\ref{45}) are consistent with the stationary points found. That is the analysis given below.
\subsection{\label{sec:sec42} Stability conditions}
For the potential to be stable, it must be bounded from below. The stability  is determined by the behavior of $V$ in the limit $K_0\rightarrow\infty$, $L_0\rightarrow\infty$ and/or $M_0\rightarrow\infty$; hence, by the signs of $J_{k(l,m)2}(\boldsymbol k,\boldsymbol l,\boldsymbol m)$ and $J_{k(l,m)4}(\boldsymbol k,\boldsymbol l,\boldsymbol m)$ in~(\ref{79}), (approach which conduces only to sufficiency conditions but not to necessary conditions).

In the strong sense, the stability of the potential is guaranteed when $V\rightarrow \infty$ for $\boldsymbol k$, $\boldsymbol l$ and $\boldsymbol m$ taking any value, which means that
\begin{equation}
\label{83}
J_{k4}(\boldsymbol k),\:J_{l4}(\boldsymbol l),\:J_{m4}(\boldsymbol m)> 0 \:\:\textrm{for all}\:\: 
|\boldsymbol k|,\:|\boldsymbol l|,\:|\boldsymbol m|\le 1.
\end{equation}
To assure the existence of a positive (semi-)definite value for $J_{k(l,m)4}(\boldsymbol k,\boldsymbol l,\boldsymbol m)$, it is sufficient to consider its value for all the stationary points of $J_{k(l,m)4}(\boldsymbol k,\boldsymbol l,\boldsymbol m)$ in the domain $|\boldsymbol k|,|\boldsymbol l|,|\boldsymbol m|<1$, and for all stationary points on the boundary $|\boldsymbol k|,|\boldsymbol l|,|\boldsymbol m|=1$. This holds, because the global minimum of the continuous function $J_{k(l,m)4}(\boldsymbol k,\boldsymbol l,\boldsymbol m)$ is reached on the compact domain $|\boldsymbol k|,|\boldsymbol l|,|\boldsymbol m|\le1$, and it is located among those stationary points. This leads to bounds on $\eta_{k(l,m)00},\eta_{k(l,m)a}$ and $\eta_{k(l,m)ab}$, which parametrise the quartic term $V_{4K(L,M)}$ of the potential. A detailed analysis of the stability criteria for a scalar potential can be found in 
Refs~\cite{b1,yith}.

With the help of Eqs.~(\ref{71a}) and~(\ref{64}), the parameters defined in~(\ref{79}), for the scalar potential in~(\ref{10a}) are:
%
{\scriptsize
\begin{equation}
\label{68aaa}
\begin{split}
 \xi_{k0}&=(\mu_1^2+\mu_2^2-\mu_3^2)/2,\:
 \eta_{k00}=(\lambda_1+\lambda_2-\lambda_3+\lambda_4)/4,
\\
\boldsymbol \xi_k&=
\begin{pmatrix}
0\\ 0\\ 0
\end{pmatrix},\:\boldsymbol \eta_k=
\begin{pmatrix}
 0\\
0\\
0
\end{pmatrix},
\\
E_k&=
\begin{pmatrix}
 (\lambda_7+2\lambda_{10})/4&0&0\\
0& \lambda_7/4&0\\
0&0&(\lambda_1+\lambda_2-\lambda_3-\lambda_4)/4
\end{pmatrix},
\\
\xi_{l0}&=(\mu_1^2-\mu_2^2+\mu_3^2)/2,\:
 \eta_{l00}=(\lambda_1-\lambda_2+\lambda_3+\lambda_5)/4,
\\
\boldsymbol \xi_l&=
\begin{pmatrix}
0\\ 0\\ 0
\end{pmatrix},\:
\boldsymbol \eta_l=
\begin{pmatrix}
 0\\
0\\
0
\end{pmatrix},
\\
 E_l&=
\begin{pmatrix}
 \lambda_8/4&0&0\\
0& \lambda_8/4&0\\
0&0&(\lambda_1-\lambda_2+\lambda_3-\lambda_5)/4
\end{pmatrix},
\\
\xi_{m0}&=(-\mu_1^2+\mu_2^2+\mu_3^2)/2,\:
 \eta_{m00}=(-\lambda_1+\lambda_2+\lambda_3+\lambda_6)/4,
\\
\boldsymbol \xi_m&=
\begin{pmatrix}
0\\ 0\\ 0
\end{pmatrix},\:
\boldsymbol \eta_m=
\begin{pmatrix}
 0\\
0\\
0
\end{pmatrix},
\\
 E_m&=
\begin{pmatrix}
 \lambda_9/4&0&0\\
0& \lambda_9/4&0\\
0&0&(-\lambda_1+\lambda_2+\lambda_3-\lambda_6)/4
\end{pmatrix}.
\end{split}
\end{equation}}
%
The stability criteria established in the previous section allow us to bound the parameters of the potential in the following way:

For $K:$
\begin{subequations}
\label{68a}
\begin{align}
\label{68aa}
 \lambda_1+\lambda_2-\lambda_3&>0,
\\
\label{68ab}
\lambda_1+\lambda_2-\lambda_3+\lambda_4&>0,
\\
\label{68ac}
\lambda_1+\lambda_2-\lambda_3+\lambda_4+\lambda_7&>0,
\\
\label{68ad}
\lambda_1+\lambda_2-\lambda_3+\lambda_4+\lambda_7+2\,\lambda_{10}&>0,
\end{align}
\end{subequations}
%

For $L:$
\begin{subequations}
\label{69a}
\begin{align}
\label{69aa}
 \lambda_1-\lambda_2+\lambda_3&>0,
\\
\label{69ab}
\lambda_1-\lambda_2+\lambda_3+\lambda_5&>0,
\\
\label{69ac}
\lambda_1-\lambda_2+\lambda_3+\lambda_5+\lambda_8&>0,
\end{align}
\end{subequations}

For $M:$
\begin{subequations}
\label{70a}
\begin{align}
\label{70aa}
 -\lambda_1+\lambda_2+\lambda_3&>0,
\\
\label{70ab}
-\lambda_1+\lambda_2+\lambda_3+\lambda_6&>0,
\\
\label{70ac}
-\lambda_1+\lambda_2+\lambda_3+\lambda_6+\lambda_9&>0.
\end{align}
\end{subequations}
In this way, when the constraints~(\ref{68a}),(\ref{69a}) and~(\ref{70a}) are satisfied, the potential is stable in the strong sense. The former constraints can be combined: summing (\ref{68aa})+(\ref{69aa}), (\ref{68aa})+(\ref{70aa}), and  (\ref{69aa})+(\ref{70aa}), we have respectively
\begin{equation}
 \label{71aa}
 \lambda_1>0,\quad
 \lambda_2>0,\quad
 \lambda_3>0.
\end{equation}
From the sums (\ref{68aa})+(\ref{69ab}), (\ref{68ab})+(\ref{69aa}), (\ref{68aa})+(\ref{70ab}),
(\ref{68ab})+(\ref{70aa}), (\ref{69aa})+(\ref{70ab}), (\ref{69ab})+(\ref{70aa}), we have
%
 \begin{equation}
\label{74aa}
\begin{split}
 2\,\lambda_1>-\lambda_5,\quad
2\,\lambda_1>-\lambda_4,\quad
2\,\lambda_2>-\lambda_6,
\\
 2\,\lambda_2>-\lambda_4,\quad
 2\,\lambda_3>-\lambda_6,\quad
 2\,\lambda_3>-\lambda_5.
\end{split}
\end{equation}
%
The following operations (\ref{68ab})+(\ref{69ab}), (\ref{68ad})+(\ref{69aa}), (\ref{68aa}) +(\ref{69ac}), (\ref{68ab})+(\ref{70ab}), (\ref{68ad})+(\ref{70aa}), (\ref{68aa})+(\ref{70ac}), (\ref{68ab})+ (\ref{70ab}), (\ref{69ac})+(\ref{70aa}),  (\ref{69aa})+(\ref{70ac}) give, respectively,
\begin{equation}
\begin{split}
2\,\lambda_1&>-(\lambda_4+\lambda_5),\quad\quad
 2\,\lambda_1>-(\lambda_4+\lambda_7),\\
2\,\lambda_1&>-(\lambda_5+\lambda_8),\quad\quad
 2\,\lambda_2>-(\lambda_4+\lambda_6),\\
 2\,\lambda_2&>-(\lambda_4+\lambda_7),\quad\quad
 2\,\lambda_2>-(\lambda_6+\lambda_9),\\
 2\,\lambda_3&>-(\lambda_5+\lambda_6),\quad\quad
 2\,\lambda_3>-(\lambda_5+\lambda_8),\\
 2\,\lambda_3&>-(\lambda_6+\lambda_9).
\end{split}
\end{equation}
Other two interesting conditions are (\ref{68ad})+(\ref{69aa}), (\ref{68ad})+ (\ref{70aa}) 
\begin{equation}
\label{76aa}
\begin{split}
2\,\lambda_1&>-(\lambda_4+\lambda_7+2\,\lambda_{10}),\\
2\,\lambda_2&>-(\lambda_4+\lambda_7+2\,\lambda_{10}).
\end{split}
\end{equation}
And else inequalities that have not yet derived here can be found.

Notice that conditions~(\ref{71aa}),~(\ref{74aa}) and~(\ref{76aa}) derived by stability conditions, are compatible with some constraints ~(\ref{mases}),~(\ref{masesp}) and~(\ref{52}) derived by positive masses~(positive concavity) conditions.

When $k_3,l_3$ and $m_3$ take fixed values, we can make use of Eqs.~(\ref{64}) and~(\ref{65aa}) in order to write the orbital variables as
{\scriptsize
\begin{equation} \label{66a}
K_0=K_0,\: L_0=K_0\left(\frac{1+k_3}{1+l_3}\right)\:\:\textrm{and}\:\;\; M_0=K_0\left(\frac{1-l_3\,k_3}{1+l_3}\right),
\end{equation}}
\hspace{-1.4mm}where $K_0$ is an independent free parameter, 
such that that the following conditions
\begin{equation}
\label{88}
 \xi_{k0}<|\boldsymbol \xi_k|,\:\textrm{and} \:\: \xi_{l0}<|\boldsymbol \xi_l|,\:\:\textrm{and} \:\:
\xi_{m0}<|\boldsymbol \xi_m|\,,
\end{equation}
imply
\pagebreak

\begin{align}
 &\left.\frac{\partial V}{\partial K_0}\right|_{
\begin{subarray}{l}
 \boldsymbol k fixed, \\ K_0=0
\end{subarray}}=\xi_{k0}+\boldsymbol\xi_k\cdot\boldsymbol k<0,\: \textrm{and}
\\
 &\left.\frac{\partial V}{\partial L_0}\right|_{
\begin{subarray}{l}
 \boldsymbol l \,fixed, \\ L_0=0
\end{subarray}}=\xi_{l0}+\boldsymbol\xi_l\cdot\boldsymbol l<0,\: \textrm{and}
\\
&\left.\frac{\partial V}{\partial M_0}\right|_{
\begin{subarray}{l}
 \boldsymbol m fixed, \\ M_0=0
\end{subarray}}=\xi_{m0}+\boldsymbol\xi_m\cdot\boldsymbol m<0.
\end{align}
for some fixed values $\boldsymbol k$, $\boldsymbol l$, and $\boldsymbol m$ while varying $K_0,\; L_0$, and $M_0$ in the form given by Eqs.~(\ref{66a}). This guarantees that the global minimum of $V$ lies at $\Phi_i\ne 0$. For our case, from~(\ref{68aaa}) and~(\ref{88}) we have that 
\[(\mu_1^2+\mu_2^2-\mu_3^2),\: (\mu_1^2-\mu_2^2+\mu_3^2),\:(-\mu_1^2+\mu_2^2+\mu_3^2)<0,\]
which implies 
\begin{equation}
\label{93a}
 \mu_1^2<0,\quad \textrm{and}\quad \mu_2^2<0,\quad\textrm{and}\quad \mu_3^2<0.
\end{equation}
%
\section{\label{sec:sec5} Stationary Points}
Now let us find the stationary points of the scalar potential, since among those points the local and global minima are located. To start with, let us define the following nine component vector
\begin{equation}
\label{84}
 \tilde{\boldsymbol P}=
\begin{pmatrix}
 K_0&
L_0&
M_0&
K_1&
K_2&
L_1&
L_2&
M_1&
M_2
\end{pmatrix}^T
\end{equation}
and let's also define the following nine vectors, each one with eight components
\begin{equation}
\label{85}
{\tilde{\boldsymbol P}}_{\{i\}}, \quad \textrm{for}\quad i=1,2,\ldots,9,
\end{equation}
where $\tilde{\boldsymbol P}_{\{i\}}$ is the vector $\tilde{\boldsymbol P}$ with the $i^{th}$ entry suppressed [for example ${\tilde{\boldsymbol P}}_{\{1\}}=\begin{pmatrix}
L_0 & M_0 & K_1 & K_2 & L_1 & L_2 & M_1 & M_2 \end{pmatrix}^T$, etc.]

With the help of this notation the potential~(\ref{10a}) reads
\begin{equation}
\label{94}
V= \tilde{\boldsymbol P}\cdot \tilde{\boldsymbol \xi}+ \tilde{\boldsymbol P}\cdot\tilde E\cdot \tilde{\boldsymbol P}
\end{equation}
where
%
{\scriptsize
\begin{equation*}
\begin{split}
\tilde{\boldsymbol \xi}=
&\begin{pmatrix}
 \frac{\mu_1^2+\mu_2^2-\mu_3^2}{2}\\
\frac{\mu_1^2-\mu_2^2+\mu_3^2}{2}\\
\frac{-\mu_1^2+\mu_2^2+\mu_3^2}{2}\\
0\\
0\\
0\\
0\\
0\\
0\\
\end{pmatrix},
%
\tilde E=
\left(\begin{matrix}
 \frac{\lambda_1+\lambda_2+\lambda_3+\lambda_4-\lambda_6-\lambda_5}{4} & \frac{\lambda_1+\lambda_6-\lambda_3-\lambda_2}{4}
\\
\frac{\lambda_1+\lambda_6-\lambda_3-\lambda_2}{4} & \frac{\lambda_1+\lambda_2+\lambda_3-\lambda_4+\lambda_5-\lambda_6}{4}
\\
\frac{\lambda_5-\lambda_3+\lambda_2-\lambda_1}{4} & \frac{\lambda_4+\lambda_3-\lambda_2-\lambda_1}{4} 
\\
0&0
\\
0&0
\\
0&0
\\
0&0
\\
0&0
\\
0&0
\end{matrix}\right.
\\
&\left.
\begin{matrix}
 \frac{\lambda_5-\lambda_3+\lambda_2-\lambda_1}{4}& 0&0&0&0&0&0
\\
 \frac{\lambda_4+\lambda_3-\lambda_2-\lambda_1}{4}& 0&0&0&0&0&0
\\
\frac{\lambda_1+\lambda_2+\lambda_3-\lambda_4-\lambda_5+\lambda_6}{4} &  0&0&0&0&0&0
\\
0& \frac{\lambda_7+2\,\lambda_{10}}{4} & 0&0&0&0&0
\\
0&0& \frac{\lambda_7}{4}&0&0&0&0
\\
0&0&0& \frac{\lambda_8}{4}&0&0&0
\\
0&0&0& 0&\frac{\lambda_8}{4}&0&0
\\
&0&0&0&0&0& \frac{\lambda_9}{4}&0
\\
&0&0&0&0&0&0& \frac{\lambda_9}{4}
\end{matrix}\right)
\end{split}
\end{equation*}}
The domain of orbital variables, Eqs.~(\ref{63}), can be written in the following form.
{\scriptsize
\begin{equation}
  \tilde{\boldsymbol P}\cdot \tilde{ g}_1\cdot  \tilde{\boldsymbol P}\ge 0,\:
\tilde{\boldsymbol P}\cdot \tilde{g}_2\cdot  \tilde{\boldsymbol P}\ge 0,\:
\tilde{\boldsymbol P}\cdot \tilde{ g}_3\cdot  \tilde{\boldsymbol P}\ge 0,\:
\: K_0\ge 0,\: L_0\ge 0,\: M_0\ge0,
\end{equation}}
with
{\footnotesize
\begin{equation}
\begin{split}
  \tilde{\boldsymbol g}_1=
\begin{pmatrix}
 1&0&0&0&0&0&0&0&0\\
0&-1&1&0&0&0&0&0&0\\
0&1&-1&0&0&0&0&0&0\\
0&0&0&-1&0&0&0&0&0\\
0&0&0&0&-1&0&0&0&0\\
0&0&0&0&0&0&0&0&0\\
0&0&0&0&0&0&0&0&0\\
0&0&0&0&0&0&0&0&0\\
0&0&0&0&0&0&0&0&0
\end{pmatrix},
\\
\tilde{\boldsymbol g}_2=
\begin{pmatrix}
 -1&0&1&0&0&0&0&0&0\\
0&1&0&0&0&0&0&0&0\\
1&0&-1&0&0&0&0&0&0\\
0&0&0&0&0&0&0&0&0\\
0&0&0&0&0&0&0&0&0\\
0&0&0&0&0&-1&0&0&0\\
0&0&0&0&0&0&-1&0&0\\
0&0&0&0&0&0&0&0&0\\
0&0&0&0&0&0&0&0&0
\end{pmatrix},
\\
\tilde{\boldsymbol g}_3=
\begin{pmatrix}
 -1&1&0&0&0&0&0&0&0\\
1&-1&0&0&0&0&0&0&0\\
0&0&1&0&0&0&0&0&0\\
0&0&0&0&0&0&0&0&0\\
0&0&0&0&0&0&0&0&0\\
0&0&0&0&0&0&0&0&0\\
0&0&0&0&0&0&0&0&0\\
0&0&0&0&0&0&0&-1&0\\
0&0&0&0&0&0&0&0&-1
\end{pmatrix}.
\end{split}
\end{equation}}
The trivial configuration $ \tilde{\boldsymbol P}=0$ is a stationary point of the potential with $V=0$, as a direct consequence of the definitions.

For the discussion of the stationary points of $V$, we must distinguish among the following cases
\begin{subequations}
\label{99a}
\begin{align}\label{99}
 \tilde{\boldsymbol P}\cdot \tilde{ g}_1\cdot  \tilde{\boldsymbol P}&> 0,\:
\tilde{\boldsymbol P}\cdot \tilde{ g}_2\cdot  \tilde{\boldsymbol P}> 0,\:
\tilde{\boldsymbol P}\cdot \tilde{\boldsymbol g}_3\cdot  \tilde{\boldsymbol P}> 0;\:
\\ \label{100}
\tilde{\boldsymbol P}\cdot \tilde{ g}_1\cdot  \tilde{\boldsymbol P}&= 0,\:
\tilde{\boldsymbol P}\cdot \tilde{ g}_2\cdot  \tilde{\boldsymbol P}> 0,\:
\tilde{\boldsymbol P}\cdot \tilde{ g}_3\cdot  \tilde{\boldsymbol P}> 0;\:
\\ \label{101}
\tilde{\boldsymbol P}\cdot \tilde{ g}_1\cdot  \tilde{\boldsymbol P}&> 0,\:
\tilde{\boldsymbol P}\cdot \tilde{ g}_2\cdot  \tilde{\boldsymbol P}= 0,\:
\tilde{\boldsymbol P}\cdot \tilde{ g}_3\cdot  \tilde{\boldsymbol P}> 0;\:
\\ \label{102}
\tilde{\boldsymbol P}\cdot \tilde{ g}_1\cdot  \tilde{\boldsymbol P}&> 0,\:
\tilde{\boldsymbol P}\cdot \tilde{ g}_2\cdot  \tilde{\boldsymbol P}> 0,\:
\tilde{\boldsymbol P}\cdot \tilde{ g}_3\cdot  \tilde{\boldsymbol P}= 0;\:
\\ \label{90e}
\tilde{\boldsymbol P}\cdot \tilde{ g}_1\cdot  \tilde{\boldsymbol P}&= 0,\:
\tilde{\boldsymbol P}\cdot \tilde{ g}_2\cdot  \tilde{\boldsymbol P}= 0,\:
\tilde{\boldsymbol P}\cdot \tilde{ g}_3\cdot  \tilde{\boldsymbol P}> 0;\:
\\ \label{99f}
\tilde{\boldsymbol P}\cdot \tilde{ g}_1\cdot  \tilde{\boldsymbol P}&= 0,\:
\tilde{\boldsymbol P}\cdot \tilde{ g}_2\cdot  \tilde{\boldsymbol P}> 0,\:
\tilde{\boldsymbol P}\cdot \tilde{ g}_3\cdot  \tilde{\boldsymbol P}= 0;\:
\\ \label{105}
\tilde{\boldsymbol P}\cdot \tilde{ g}_1\cdot  \tilde{\boldsymbol P}&> 0,\:
\tilde{\boldsymbol P}\cdot \tilde{ g}_2\cdot  \tilde{\boldsymbol P}= 0,\:
\tilde{\boldsymbol P}\cdot \tilde{ g}_3\cdot  \tilde{\boldsymbol P}= 0;\:
\\ \label{106}
\tilde{\boldsymbol P}\cdot \tilde{ g}_1\cdot  \tilde{\boldsymbol P}&= 0,\:
\tilde{\boldsymbol P}\cdot \tilde{ g}_2\cdot  \tilde{\boldsymbol P}= 0,\:
\tilde{\boldsymbol P}\cdot \tilde{ g}_3\cdot  \tilde{\boldsymbol P}= 0.\:
\end{align}
\end{subequations}
The stationary points of $V$ in the inner part of the domain, cases~(\ref{99}),~(\ref{101}),~(\ref{102}) and~(\ref{105}), imply linear independence between $\langle\Phi_1\rangle$ and $\langle\Phi_2\rangle$, something which is not allowed. The global minimum for the case~(\ref{106}) implies LD among $\langle\Phi_1\rangle$, $\langle\Phi_2\rangle$ and $\langle\Phi_3\rangle$, not allowed either. Case~(\ref{90e}) implies $\langle\Phi_1\rangle=0$. So, the only two cases of concern for us here are (\ref{100}) corresponding to the general vacuum structure with $v_1,v_2,v_3,V_1,V_2\neq 0$ studied in Sect.~(\ref{sec:sec321}), and case  (\ref{99f}) which corresponds to the vacuum structure $\langle\Phi_2\rangle=0$ studied in Sect.~(\ref{sec:sec322}).


In general, the stationary points of the scalar potential in~(\ref{94}), for any domain in~(\ref{99a}), are stationary points of the function
\begin{equation}
 \tilde F(\tilde{\boldsymbol P},u,v,w)=V-u\:\tilde{\boldsymbol P}\cdot\tilde g_1\cdot\tilde{\boldsymbol P}-
v\:\tilde{\boldsymbol P}\cdot\tilde g_2\cdot\tilde{\boldsymbol P}-
w\:\tilde{\boldsymbol P}\cdot\tilde g_3\cdot\tilde{\boldsymbol P},
\end{equation}
where $u,v$ and $w$ are Lagrange multipliers. The relevant stationary points of $\tilde F$ are thus given as solutions to the equation

\begin{subequations}
\label{100d}
\begin{align}
\label{100a}
 &(\tilde E-u\tilde g_1-v\tilde g_2-w\tilde g_3)\tilde{\boldsymbol P}=-\frac{1}{2}\tilde{\boldsymbol \xi};\quad
\textrm{with}
\\
\label{101c}
&\tilde{\boldsymbol P}\cdot\tilde g_1\cdot\tilde{\boldsymbol P}\ge0,\:
\tilde{\boldsymbol P}\cdot\tilde g_2\cdot\tilde{\boldsymbol P}\ge0,\:
\tilde{\boldsymbol P}\cdot\tilde g_3\cdot\tilde{\boldsymbol P}\ge0,
\end{align}
\end{subequations}
and
\begin{equation}\label{93c}
K_0,L_0,M_0>0,
\end{equation}
with the inequality ($>0$) in Eq.~(\ref{101c}) taking place, for the case when the Lagrange multipliers are excluded. For regular values of $u,v$ and $w$, with the determinant
\[\det(\tilde E-u\tilde g_1-v\tilde g_2-w\tilde g_3)\ne0\] 
we have
\begin{equation}
\label{101e}
 \tilde{\boldsymbol P}=-\frac{1}{2}(\tilde E-u\tilde g_1-v\tilde g_2-w\tilde g_3)^{-1}\tilde{\boldsymbol \xi}.
\end{equation}
The Lagrange multipliers are thus obtained by inserting~(\ref{101e}) in the constraint Eqs.~(\ref{101c}):
%
{\scriptsize
\begin{equation}
\begin{split}
\tilde{\boldsymbol \xi}(\tilde E-u\tilde g_1-v\tilde g_2-w\tilde g_3)^{-1}&\tilde g_1(\tilde E-u\tilde g_1-v\tilde g_2-w\tilde g_3)^{-1} \tilde{\boldsymbol \xi}=0,
\\
\tilde{\boldsymbol \xi}(\tilde E-u\tilde g_1-v\tilde g_2-w\tilde g_3)^{-1}&\tilde g_2(\tilde E-u\tilde g_1-v\tilde g_2-w\tilde g_3)^{-1} \tilde{\boldsymbol \xi}=0,
\\
\tilde{\boldsymbol \xi}(\tilde E-u\tilde g_1-v\tilde g_2-w\tilde g_3)^{-1}&\tilde g_3(\tilde E-u\tilde g_1-v\tilde g_2-w\tilde g_3)^{-1} \tilde{\boldsymbol \xi}=0,
\\
\textrm{and}\quad &K_0,L_0,M_0>0.
\end{split}
\end{equation}}
%
\hspace{-1.0mm}Additionally, there may be up to 9 values $u=\tilde\mu_a$ (and also 9 values for $v=\tilde\mu_a$, and for  $w=\tilde\mu_a$) with $a=1,\ldots,9$, for which $\det(\tilde E-u\tilde g_1-v\tilde g_2-w\tilde g_3)=0$. Depending on the form of the potential; some, or all of them, may lead to exceptional solutions of~(\ref{100a}).

For \textit{any}  stationary point of the potential we have
\begin{equation}
\label{103}
\left.V\right|_{stat}=\frac{1}{2}\tilde{\boldsymbol P}\cdot\tilde{\boldsymbol\xi}=-\tilde{\boldsymbol P}\cdot\tilde E\cdot\tilde{\boldsymbol P}.
\end{equation}
Suppose now that the strong stability condition~(\ref{83}) holds. Then~(\ref{103}) gives for non-trivial stationary points where $\tilde{\boldsymbol P}\ne0$:
\begin{equation}
 \left.V\right|_{stat}<0.
\end{equation}
Firstly, 
in the context that only one Lagrange multiplier is non-zero, for instance $u_p\neq0, v=w=0$ in~(\ref{100a}),  let's consider $\tilde{\boldsymbol p}=(p_{k_0},p_{l_0},p_{m_0},p_{k_1},p_{k_2},p_{l_1},p_{l_2},p_{m_1},p_{m_2})^T$ be a stationary point for this case. Then, from~(\ref{94}) and~(\ref{100a}) we have
\begin{subequations}
\label{97}
\begin{align}
\label{105aa}
\left.\frac{\partial V}{\partial K_0}\right|_{
\begin{subarray}
\:{\tilde{\boldsymbol P}_{\{1\}}}\: fixed,\\
\tilde{\boldsymbol P}=\tilde{\boldsymbol p}
\end{subarray}
}&=2\:u_p\:p_{k_0},
\\
\left.\frac{\partial V}{\partial L_0}\right|_{
\begin{subarray}
\:{\tilde{\boldsymbol P}_{\{2\}}}\: fixed,\\
\tilde{\boldsymbol P}=\tilde{\boldsymbol p}
\end{subarray}
}&=2\:u_p\:p_{l_0},
\\
\left.\frac{\partial V}{\partial M_0}\right|_{
\begin{subarray}
\:{\tilde{\boldsymbol P}_{\{3\}}}\: fixed,\\
\tilde{\boldsymbol P}=\tilde{\boldsymbol p}
\end{subarray}
}&=2\:u_p\:p_{m_0},
\end{align}
\end{subequations}
where the notation established in~(\ref{85}) has been used. For the analysis which follows, only the most convenient partial derivative from~(\ref{97}) is chosen, in such a way that only two, out of the three values in~(\ref{64}) lower down, and also that the inequalities~(\ref{63}) hold for the new points; (for example, let us take $M_0$. From~(\ref{64}) we have ($K_3,L_3)\rightarrow (K^\prime_3=K_3-\Delta M_0),(L^\prime_3=L_3-\Delta M_0)$. Where $|K_3^\prime|<|K_3|$ and $|L_3^\prime|<|L_3|$, if $K_3,L_3>0$). If $u_p<0$, there are points $\tilde{\boldsymbol P}$ with $K_0>p_{k_0}$ (or $L_0>p_{l_0},M_0>p_{m_0})$, $\tilde{\boldsymbol P}_{\{1\}}=\tilde{\boldsymbol p}_{\{1\}}$ (or $\tilde{\boldsymbol P}_{\{2\}}=\tilde{\boldsymbol p}_{\{2\}}$, $\tilde{\boldsymbol P}_{\{3\}}=\tilde{\boldsymbol p}_{\{3\}}$) for which the potential decreases in its neighborhood, and as a consequence cannot be a minimum.
We conclude that in a theory with the required EWSB,  a stationary point coming from an unique no null Lagrange multiplier (for example $u_0\neq0, v=w=0$), to be a global minimum candidate, it must hold
\begin{equation}
\label{99aa}
 u_0>0.
\end{equation}
Secondly, for the stationary points $\tilde{\boldsymbol p}$ and $\tilde{\boldsymbol q}$, we have in general, from (\ref{100a}) and~(\ref{103}), the following relation
{\footnotesize
\begin{equation}
\label{106f}
\begin{split}
V(\tilde{\boldsymbol p})-V(\tilde{\boldsymbol q})&=
\frac{1}{2}\tilde{\boldsymbol p}\cdot\tilde{\boldsymbol{\xi}}
-\frac{1}{2}\tilde{\boldsymbol q}\cdot\tilde{\boldsymbol{\xi}}
\\
&=\tilde{\boldsymbol p}\cdot(u_q\tilde g_1+v_q\tilde g_2+w_q\tilde g_3-\tilde E)\cdot\tilde{\boldsymbol q}
\\
&-\tilde{\boldsymbol q}\cdot(u_p\tilde g_1+v_p\tilde g_2+w_p\tilde g_3-\tilde E)\cdot\tilde{\boldsymbol p}
\\
&=
(u_q-u_p)\:\tilde{\boldsymbol p}\cdot\tilde g_1\cdot\tilde{\boldsymbol q}+
(v_q-v_p)\:\tilde{\boldsymbol p}\cdot\tilde g_2\cdot\tilde{\boldsymbol q}
\\
&+(w_q-w_p)\:\tilde{\boldsymbol p}\cdot\tilde g_3\cdot\tilde{\boldsymbol q},
\end{split}
\end{equation}}
%
\hspace{-1.4mm}where $\tilde{\boldsymbol p}$ and $\tilde{\boldsymbol q}$ are vectors on the forward light cone, and $\tilde{\boldsymbol p}\cdot\tilde g_1\cdot\tilde{\boldsymbol q}$, $\tilde{\boldsymbol p}\cdot\tilde g_2\cdot\tilde{\boldsymbol q}$,
$\tilde{\boldsymbol p}\cdot\tilde g_3\cdot\tilde{\boldsymbol q}$ are always non-negative.

In Table~\ref{c3},  an exhaustive of all the possible stationary
points of the potential are presented~(even if it includes unphysical
VEVs). Lagrange multipliers coming from solutions of Eq.~(\ref{100a}) belonging to regulars and exceptional values for the cases are stated. Those cases of Lagrange multipliers giving the same stationary point, only one of them were included in Table~\ref{c3}.  Solutions which imply specific relations among the parameters $\mu_1^2,\mu_2^2, \mu_3^2$ were excluded too~(see for example~\cite{yith}). Incidentally, it is important to say that all stationary points written in Table~\ref{c3} give physical VEVs, i.e, they are consistent in relation with matrices~(\ref{45}).

\renewcommand{\arraystretch}{0.5}
\begin{table*}
\centering
\begin{tabular}{|
c|c|c|}\hline
& & \\[-1.3mm]
 Domains & Lagrange Multipliers & Stationary Points\\ & & \\[-3.3mm]\\ \hline \hline&&\\ [-1.3mm]
\multirow{2}{*}
{\footnotesize$
{
\begin{matrix}
 \tilde{\boldsymbol P}_{1}\cdot\tilde g_1\cdot\tilde{\boldsymbol P}_1=0,\\
\\
 \tilde{\boldsymbol P}_1\cdot\tilde g_2\cdot \tilde{\boldsymbol P}_1=4\,K_{01}\,M_{01},\\
\\
\tilde{\boldsymbol P}_1\cdot\tilde g_3\cdot\tilde{\boldsymbol P}_1=0.
\end{matrix}}
$}
&
{\footnotesize
$u_1= \frac{ {\mu_2}^{2}+\lambda_4\, K_{01} +\lambda_6\,M_{01}}{4\,K_{01}},\quad w_1=0$}&
{\scriptsize
$\underline{\tilde{\boldsymbol P}}_1=
\begin{pmatrix}
 K_{01}=\frac{\lambda_5\mu_3^2-2\lambda_3\mu_1^2}{4\lambda_1\lambda_3-\lambda_5^2}
\\
L_{01}=K_{01}+M_{01}
\\
M_{01}=\frac{\lambda_5\mu_1^2-2\lambda_1\mu_3^2}{4\lambda_1\lambda_3-\lambda_5^2}
\end{pmatrix}
$}
\\ [6mm]
\cline{2-3}&&\\[-1.1mm]
&{\footnotesize$u_2=0,\quad w_2=\frac{{\mu_2}^{2} +\lambda_4\,K_{01} +\lambda_6\,M_{01}}{4\,M_{01}}$}
&
${\tilde{\boldsymbol P}_2}={\tilde{\boldsymbol P}_1}$\\[2.0mm]
\hline
&&\\[-1.3mm]
{\footnotesize 
$
{
\begin{matrix}
 \tilde{\boldsymbol P}_3\cdot\tilde g_1\cdot\tilde{\boldsymbol P}_3=0,\\
 \tilde{\boldsymbol P}_3\cdot\tilde g_2\cdot \tilde{\boldsymbol P}_3=0,\\
\tilde{\boldsymbol P}_3\cdot\tilde g_3\cdot\tilde{\boldsymbol P}_3=4\,K_{03}\,L_{03}.
\end{matrix}}
$}
&{\footnotesize
$u_3=\frac{{\mu_1}^{2}+\lambda_4\,K_{03} +\lambda_5\,L_{03}}{4\,K_{03}},\quad v_3=0$}&
{\scriptsize
$\underline{\tilde{\boldsymbol P}_3}=
\begin{pmatrix}K_{03}= \frac{\lambda_6\mu_3^2-2\lambda_3\mu_2^2}{4\lambda_2\lambda_3-\lambda_6^2}
\\
L_{03}=\frac{\lambda_6\mu_2^2-2\lambda_2\mu_3^2}
{4\lambda_2\lambda_3-\lambda_6^2}
\\
K_{03}+L_{03}
\end{pmatrix}
$}
\\ [7mm] 
 \hline
&&\\[-1.0mm]
{\footnotesize
$
{
\begin{matrix}
 \tilde{\boldsymbol P}_4\cdot\tilde g_1\cdot\tilde{\boldsymbol P}_4=4\,L_4\,M_4,\\
 \tilde{\boldsymbol P}_4\cdot\tilde g_2\cdot \tilde{\boldsymbol P}_4=0,\\
\tilde{\boldsymbol P}_4\cdot\tilde g_3\cdot\tilde{\boldsymbol P}_4=0.
\end{matrix}}
$}
&{\footnotesize
$v_4=0,\quad w_4= \frac{{\mu_3}^{2}+\lambda_5\, L_4+\lambda_6\,M_4}{4\,M_4}$}
&
{\scriptsize
$\underline{\tilde{\boldsymbol P}_4}=
\begin{pmatrix}
L_4+M_4
\\ 
L_4=\frac{\lambda_4\mu_2^2-2\lambda_2\mu_1^2}
{4\lambda_1\lambda_2-\lambda_4^2}
\\
M_4=\frac{\lambda_4\mu_1^2-2\lambda_1\mu_2^2}{4\lambda_1\lambda_2-\lambda_4^2}
\end{pmatrix}
$}
\\ [6.5mm]
\hline &&\\[-1.0mm]
\multirow{2}{*}
{\footnotesize 
$
\begin{matrix}
 \tilde{\boldsymbol P}_{\{5,6\}}\cdot\tilde g_1\cdot\tilde{\boldsymbol P}_{\{5,6\}}=0,\\
 \tilde{\boldsymbol P}_{\{5,6\}}\cdot\tilde g_2\cdot \tilde{\boldsymbol P}_{\{5,6\}}\neq0,\\
\tilde{\boldsymbol P}_{\{5,6\}}\cdot\tilde g_3\cdot\tilde{\boldsymbol P}_{\{5,6\}}\neq0.
\end{matrix}
$}
& Exceptional solution:\:\: $u_5=-\frac{\lambda_7+2\:\lambda_{10}}{4}$&
$\tilde{\boldsymbol P}_5$
\\ [3mm] 
\cline{2-3}
&&\\[-0.2mm]
& Exceptional solution:\:\: $u_6=-\frac{\lambda_7}{4}$&
$\tilde{\boldsymbol P}_6$
\\ [2.8mm]
\hline &&\\[-1.3mm]
{\footnotesize
$
\begin{matrix}
\tilde{\boldsymbol P}_7\cdot\tilde g_1\cdot\tilde{\boldsymbol P}_7\neq0,\\
 \tilde{\boldsymbol P}_7\cdot\tilde g_2\cdot \tilde{\boldsymbol P}_7=0,\\
\tilde{\boldsymbol P}_7\cdot\tilde g_3\cdot\tilde{\boldsymbol P}_7\neq0.
\end{matrix}
$}
& Exceptional solution:\:\: $v_7=-\frac{\lambda_8}{4}$&
$\tilde{\boldsymbol P}_7$
\\ [4.5mm]
 \hline &&\\[-1.0mm]
{\footnotesize
$
\begin{matrix}
 \tilde{\boldsymbol P}_8\cdot\tilde g_1\cdot\tilde{\boldsymbol P}_8\neq0,\\
 \tilde{\boldsymbol P}_8\cdot\tilde g_2\cdot \tilde{\boldsymbol P}_8\neq0,\\
\tilde{\boldsymbol P}_8\cdot\tilde g_3\cdot\tilde{\boldsymbol P}_8=0.
\end{matrix}
$}
& Exceptional solution:\:\: $w_8=-\frac{\lambda_9}{4}$&
$\tilde{\boldsymbol P}_8
$
\\[4.5mm] \hline &&  \\[-1mm]
{\footnotesize
$
\begin{matrix}
 \tilde{\boldsymbol P}_9\cdot\tilde g_1\cdot\tilde{\boldsymbol P}_9=0,\\
 \tilde{\boldsymbol P}_9\cdot\tilde g_2\cdot \tilde{\boldsymbol P}_9=0,\\
\tilde{\boldsymbol P}_9\cdot\tilde g_3\cdot\tilde{\boldsymbol P}_9=0.
\end{matrix}
$}
& 
{\footnotesize
$\begin{matrix}
\textrm{Exceptional solution:}\:\:\\
 u_9=-\frac{\lambda_7+2\,\lambda_{10}}{4},\\
v_9=\frac{\mu_3^2+\lambda_5\,L_{9}+\lambda_6\,M_{9}}{4\,L_{9}},\quad w_9=0
%
\end{matrix}
$
}&
{\scriptsize
$\underline{\tilde{\boldsymbol P}_9}=
\begin{pmatrix}
L_{9}+M_{9}
\\
L_{9}=\frac{\left(\lambda_4+\lambda_7+2\,\lambda_{10}\right)\,{\mu_2}^{2}-2\,\lambda_2\,{\mu_1}^{2}}{4\,\lambda_1\,\lambda_2-\left( \lambda_4+\lambda_7+2\,\lambda_{10}\right)^{2}}
\\
M_{9}=\frac{\left(\lambda_4+\lambda_7+2\,\lambda_{10}\right)\,{\mu_1}^{2}-2\,\lambda_1\,{\mu_2}^{2}}{4\,\lambda_1\,\lambda_2-\left( \lambda_4+\lambda_7+2\,\lambda_{10}\right)^{2}}
\\
\pm2\sqrt{L_{9}\:M_{9}}
\end{pmatrix}
$}
\\[7mm] \hline &&  \\[-1mm]
{\footnotesize
$
\begin{matrix}
 \tilde{\boldsymbol P}_{10}\cdot\tilde g_1\cdot\tilde{\boldsymbol P}_{10}=0,\\
 \tilde{\boldsymbol P}_{10}\cdot\tilde g_2\cdot \tilde{\boldsymbol P}_{10}=0,\\
\tilde{\boldsymbol P}_{10}\cdot\tilde g_3\cdot\tilde{\boldsymbol P}_{10}=0.
\end{matrix}
$}
& 
{\footnotesize
$\begin{matrix}
\textrm{Exceptional solution:}\\
 u_{10}=-\frac{\lambda_7}{4},\\
 v_{10}=\frac{\mu_3^2+\lambda_5\,L_{10}+\lambda_6\,M_{10}}{4\,L_{10}},
\quad w_{10}=0
\end{matrix}
$}&
{\scriptsize
$\underline{\tilde{\boldsymbol P}_{10}}=
\begin{pmatrix}
L_{10}+M_{10}
\\
L_{10}=\frac{\left(\lambda_4+\lambda_7\right)\,{\mu_2}^{2}-2\,\lambda_2\,{\mu_1}^{2}}{{4\,\lambda_1\,\lambda_2-\left( \lambda_4+\lambda_7\right)}^{2}}
\\
M_{10}=\frac{\left(\lambda_4+\lambda_7\right)\,{\mu_1}^{2}-2\,\lambda_1\,{\mu_2}^{2}}{4\,\lambda_1\,\lambda_2-{\left( \lambda_4+\lambda_7\right)}^{2}}
\\
0
\\
\pm2\sqrt{L_{10}\:M_{10}}
\end{pmatrix}
$}
\\[7.5mm] \hline &&  \\[-1mm]
{\footnotesize
$
\begin{matrix}
 \tilde{\boldsymbol P}_{11}\cdot\tilde g_1\cdot\tilde{\boldsymbol P}_{11}=0,\\
 \tilde{\boldsymbol P}_{11}\cdot\tilde g_2\cdot \tilde{\boldsymbol P}_{11}=0,\\
\tilde{\boldsymbol P}_{11}\cdot\tilde g_3\cdot\tilde{\boldsymbol P}_{11}=0.
\end{matrix}
$}
& 
{\footnotesize
$\begin{matrix}
\textrm{Exceptional solution:}\\
u_{11}=\frac{\mu_2^2+\lambda_4\,K_{11}+\lambda_6\,M_{11}}{4\,K_{11}},\\
  v_{11}=-\frac{\lambda_8}{4},\quad w_{11}=0
\end{matrix}
$}&
{\scriptsize
$\underline{\tilde{\boldsymbol P}_{11}}=
\begin{pmatrix}
K_{11}=\frac{\left(\lambda_5+\lambda_8\right)\,{\mu_3}^{2}-2\,\lambda_3\,{\mu_1}^{2}}{4\,\lambda_1\,\lambda_3-{\left( \lambda_5+\lambda_8\right)}^{2}}
\\
K_{11}+M_{11}
\\
M_{11}=\frac{\left(\lambda_5+\lambda_8\right)\,{\mu_1}^{2}-2\,\lambda_1\,{\mu_3}^{2}}{4\,\lambda_1\,\lambda_3-{\left( \lambda_5+\lambda_8\right)}^{2}}
\\
0
\\
0
\\
\pm2\sqrt{K_{11}\:M_{11}}
\end{pmatrix}
$}
\\[7.5mm] \hline &&  \\[-1.mm]
{\footnotesize
$
\begin{matrix}
 \tilde{\boldsymbol P}_{12}\cdot\tilde g_1\cdot\tilde{\boldsymbol P}_{12}=0,\\
 \tilde{\boldsymbol P}_{12}\cdot\tilde g_2\cdot \tilde{\boldsymbol P}_{12}=0,\\
\tilde{\boldsymbol P}_{12}\cdot\tilde g_3\cdot\tilde{\boldsymbol P}_{12}=0.
\end{matrix}
$}
& 
{\footnotesize
$\begin{matrix}
\textrm{Exceptional solution:}\\
u_{12}=\frac{\mu_1^2+\lambda_4\,K_{12}+\lambda_5\,L_{12}}{4\,K_{12}},\\
v_{12}=-\frac{\lambda_9}{4},\quad w_{12}=0
\end{matrix}
$}&
{\scriptsize
$\underline{\tilde{\boldsymbol P}_{12}}=
\begin{pmatrix}
K_{12}=\frac{\left(\lambda_6+\lambda_9\right)\,{\mu_3}^{2}-2\,\lambda_3\,{\mu_2}^{2}}{4\,\lambda_2\,\lambda_3-{\left( \lambda_6+\lambda_9\right)}^{2}}
\\
L_{12}=\frac{\left(\lambda_6+\lambda_9\right)\,{\mu_2}^{2}-2\,\lambda_2\,{\mu_3}^{2}}{4\,\lambda_2\,\lambda_3-{\left( \lambda_6+\lambda_9\right)}^{2}}
\\
K_{12}+L_{12}
\\
0
\\
0
\\
0
\\
0
\\
\pm2\sqrt{K_{12}\:L_{12}}
\end{pmatrix}
$}
\\[10.5mm] \hline &&  \\[-1mm]
\multirow{14}{*}
{{\footnotesize
$
\begin{matrix}
 \tilde{\boldsymbol P}\cdot\tilde g_1\cdot\tilde{\boldsymbol P}=0,\\
 \tilde{\boldsymbol P}\cdot\tilde g_2\cdot \tilde{\boldsymbol P}=0,\\
\tilde{\boldsymbol P}\cdot\tilde g_3\cdot\tilde{\boldsymbol P}=0.
\end{matrix}
$}}
& $u_{13}=\frac{\lambda_4\,\mu_1^2-2\,\lambda_1\,\mu_2^2}{4\,\mu_1^2},\:\:  v_{13}=\frac{\lambda_5\,\mu_1^2-2\,\lambda_1\,\mu_3^2}{4\,\mu_1^2},\:\: w_{13}=0$&
{\scriptsize$\underline{\tilde{\boldsymbol P}_{13}}=
\begin{pmatrix}
 -\frac{\mu_1^2}{2\lambda_1}
\\
-\frac{\mu_1^2}{2\lambda_1}
\end{pmatrix}$}
\\ [4.0mm]
 \cline{2-3} &&\\[-1mm]
& $u_{14}=\frac{\lambda_4\,\mu_2^2-2\,\lambda_2\,\mu_1^2}{4\,\mu_2^2},\:\:v_{14}=0,\:\: w_{14}=
\frac{\lambda_6\,\mu_2^2-2\,\lambda_2\,\mu_3^2}{4\,\mu_2^2}$&
{\scriptsize
$\underline{\tilde{\boldsymbol P}_{14}}=
\begin{pmatrix}
 -\frac{\mu_2^2}{2\lambda_2}
\\
0
\\
-\frac{\mu_2^2}{2\lambda_2}
\end{pmatrix}$}
\\ [5.5mm]
\cline{2-3}  &&\\[-1mm]
& $u_{15}=0,\:\: v_{15}=\frac{\lambda_5\,\mu_3^2-2\,\lambda_3\,\mu_1^2}{4\,\mu_3^2},\:\:  w_{15}=
\frac{\lambda_6\,\mu_3^2-2\,\lambda_3\,\mu_2^2}{4\,\mu_3^2}$&
{\scriptsize
$\underline{\tilde{\boldsymbol P}_{15}}=
\begin{pmatrix}
0
\\
 -\frac{\mu_3^2}{2\lambda_3}
\\
-\frac{\mu_3^2}{2\lambda_3}
\end{pmatrix}$}
\\  [5.5mm]\hline
\end{tabular}
\caption{Lagrange Multipliers. The stationary points ($\underline{\tilde{\boldsymbol P}}$) were underlined indicating  that only the upper non-zero entries of column vector are written explicitly, with the remaining ones entries filled by zeros.}
\label{c3}
\end{table*}
To end our analysis, let us apply our findings to the two independent vacuum structures given by the constraints (\ref{100}) and~(\ref{99f}), which were studied in detail in Sects. (\ref{sec:sec321}) and~(\ref{sec:sec322}) respectively.

\subsection{\label{sec:sec51} Case $v_2=V_2=0$}
We want a global minimum with the configuration~(\ref{53a}), which implies solutions satisfying~(\ref{99f}). For this purpose we use the results in Table~(\ref{c3}), the stability conditions stated in Sec.~(\ref{sec:sec42}) and taking into account expression~(\ref{106f}).
 In the case of an unique no null Lagrange multiplier, the restriction in~(\ref{99aa}) can be used. In that way, the conditions found below are sufficient (but not necessary) to have the minimum of the scalar potential~(\ref{10a}) at $\tilde{\boldsymbol P}_1$. 

From the Table~\ref{c3},  we want the global minimum be associated to Lagrange multiplier $u_1$(  where $w_2$ gives the same stationary point) which does not coincide with solutions inside the forward light cone~(\ref{99}). 

Let's assume the following conditions
\begin{equation}
\label{102aa}
 \lambda_4>0,\quad\lambda_5<0\quad\textrm{and}\quad\lambda_6>0,
\end{equation}
which in combination with the inequalities~(\ref{71aa}),~(\ref{74aa}) and~(\ref{93a}), we have that $\lambda_5\mu_3^2-2\lambda_3\mu_1^2>0$, $\lambda_5\mu_1^2-2\lambda_1\mu_3^2>0$, and $4\lambda_1\lambda_3-\lambda_5^2>0$, hence
%
$ K_{01}>0, M_{01}>0$ and $L_{01}>0$.
%
Additionally, with same arguments, we can verify that the Lagrange multipliers $v_{13},v_{15}<0$.

As you can see, there is not inconvenient to impose the following condition
\begin{equation}
\label{101a}
 u_1>0,
\end{equation}
as is required by the global minimum condition~(\ref{99aa}).
Therefore, for the moment, the point $\tilde{\boldsymbol P}_1$ satisfy all requirements to be a stationary point. The other aspect to take into account, it is to show that this point is the global minimum of potential. For that, let's see the other Lagrange multipliers and their points, and to establish conditions over them such that the global are not found  there.

For example, if we assume
\begin{equation}
\label{108}
\lambda_7+2\lambda_{10}>0,\quad\lambda_7>0,\quad \lambda_8>0\quad\lambda_9>0,
\end{equation}
it immediately discards out the points $\tilde{\boldsymbol P}_5,\tilde{\boldsymbol P}_6,\tilde{\boldsymbol P}_7$ and $\tilde{\boldsymbol P_8}$ as minimal global points of potential, because the corresponding  Lagrange multipliers $u_5,u_6,v_7$ and $w_8$ are negative numbers.

If we assume
\begin{align}
\label{106aa}
 \lambda_6\mu_2^2-2\lambda_2\mu_3^2>0,\\
\label{107aa}
\lambda_6\mu_3^2-2\lambda_3\mu_2^2<0,
\end{align}
it gives either the condition $K_{03}>0$ or $L_{03}>0$, but not both conditions satisfied simultaneously. And, in similar way, considering the case
\begin{align}
\label{108aa}
 \lambda_4\mu_2^2-2\lambda_2\mu_1^2>0,\\
\label{109aa}
\lambda_4\mu_1^2-2\lambda_1\mu_2^2<0
\end{align}
we conclude that $L_4,M_4$ are not positive numbers simultaneously. Then, $\tilde{\boldsymbol 
P_3}$ and $\tilde{\boldsymbol P_4}$ are not stationary points of potential.

From~(\ref{93a}),~(\ref{108aa}) and~(\ref{106aa}) we see that the Lagrange multipliers $u_{14},w_{14}<0$ . Thus, the minimum of potential is not present at $\tilde{\boldsymbol P}_{14}$.

We derived above that $v_{13}<0$ and $v_{15}<0$, that together with  conditions~(\ref{93a}) and~(\ref{102aa}), 
it is easy to verify that $u_1-u_{13}=\left[v_{13}(\lambda_6\mu_1^2-\lambda_5\mu_2^2)\right]/(4\mu_3^2v_{15})>0$, which implies that $u_1>u_{13}$. In the same way, $w_2-w_{15}=\left[v_{15}(\lambda_4\mu_3^2-\lambda_5\mu_2^2)\right]/(4\mu_1^2v_{13})>0$, that is, $w_2>w_{15}$. Therefore, in the points $\tilde{\boldsymbol P}_{13}$ and $\tilde{\boldsymbol P}_{15}$ the global minimum are not found.

Remain to see the points $\tilde{\boldsymbol P}_{9},\tilde{\boldsymbol P}_{10},\tilde{\boldsymbol P}_{11}$ and $\tilde{\boldsymbol P}_{12}$. In order to discard these points as global minima, we can proceed in the same way as we did with the points $\tilde{\boldsymbol P}_{3}$ and $\tilde{\boldsymbol P}_{4}$. Let's consider the numerator of $L_9,L_{10},K_{11},L_{12}$ as positive and the numerator of $M_9,M_{10},M_{11},K_{12}$ as negative. After that, we obtain  the following conditions,
\begin{align}
 \frac{2\lambda_1\mu_2^2}{\mu_1^2}&<\lambda_4<\frac{2\lambda_2\mu_1^2}{\mu_2^2}-\max\left\{\lambda_7,(\lambda_7+2\lambda_{10})\right\},
\\
\frac{2\lambda_3\mu_2^2}{\mu_3^2}&<\lambda_6<\frac{2\lambda_2\mu_3^2}{\mu_2^2}-\lambda_9,
\\
\frac{2\lambda_1\mu_3^2}{\mu_1^2}&<(\lambda_5+\lambda_8)<\frac{2\lambda_3\mu_1^2}{\mu_3^2},
\end{align}
where the inequalities~(\ref{106aa}) to~(\ref{109aa}) are derived from these new ones. And where the function $\max$ takes the largest value from a set.

Finally, we conclude under conditions above, the global minimum of the potential lies on  point $\tilde{\boldsymbol P}_1$, where
\begin{equation}
 \tilde{\boldsymbol P}_1\cdot\tilde g_2\cdot \tilde{\boldsymbol P}_1=4\;K_{01}\;M_{01}>0.
\end{equation}
Also
{\footnotesize
\begin{align}
 \langle\underline K\rangle&=
\begin{pmatrix}
 \langle\Phi_1^\dag\Phi_1\rangle& \langle\Phi_2^\dag\Phi_1\rangle\\
\langle\Phi_1^\dag\Phi_2\rangle& \langle\Phi_2^\dag\Phi_2\rangle
\end{pmatrix}=
\begin{pmatrix}
 \frac{\lambda_5\mu_3^2-2\lambda_3\mu_1^2}{4\lambda_1\lambda_3-\lambda_5^2}& 0\\
0&0
\end{pmatrix},
\\
\langle\underline L\rangle&=
\begin{pmatrix}
 \langle\Phi_1^\dag\Phi_1\rangle& \langle\Phi_3^\dag\Phi_1\rangle\\
\langle\Phi_1^\dag\Phi_3\rangle& \langle\Phi_3^\dag\Phi_3\rangle
\end{pmatrix}=
\begin{pmatrix}
 \frac{\lambda_5\mu_3^2-2\lambda_3\mu_1^2}{4\lambda_1\lambda_3-\lambda_5^2}& 0\\
0&\frac{\lambda_5\mu_1^2-2\lambda_1\mu_3^2}{4\lambda_1\lambda_3-\lambda_5^2}
\end{pmatrix},
\\
\langle\underline M\rangle&=
\begin{pmatrix}
 \langle\Phi_2^\dag\Phi_2\rangle& \langle\Phi_3^\dag\Phi_2\rangle\\
\langle\Phi_2^\dag\Phi_3\rangle& \langle\Phi_3^\dag\Phi_3\rangle
\end{pmatrix}=
\begin{pmatrix}
 0& 0\\
0&\frac{\lambda_5\mu_1^2-2\lambda_1\mu_3^2}{4\lambda_1\lambda_3-\lambda_5^2}
\end{pmatrix},
\end{align}}
\hspace{-1.7mm}which implies the configuration of VEVs vectors given by~(\ref{53a}). Let us mention here that the former results does not change if we consider complex phases in the VEV. Then we have
\begin{align}
\label{115b}
 \frac{v_1^2+V_1^2}{2}&=\frac{\lambda_5\mu_3^2-2\lambda_3\mu_1^2}{4\lambda_1\lambda_3-\lambda_5^2},
\\
\frac{v_3^2}{2}&=\frac{\lambda_5\mu_1^2-2\lambda_1\mu_3^2}{4\lambda_1\lambda_3-\lambda_5^2},
\end{align}
solutions that agree with the constraint equations given in~(\ref{54}). Using the former relations we can write the Lagrange multiplier $u_1$ in the following way
{\scriptsize
\begin{equation}
\label{117}
 u_1=\frac{1}{4}\frac{(4\lambda_1\lambda_3-\lambda_5^2)}{(\lambda_5\mu_3^2-2\lambda_3\mu_1^2)}
\left(\frac{\lambda_4\,(v_1^2+V_1^2)+\lambda_6\,v_3^2}{2}+\mu_2^2\right)>0.
\end{equation}}
\hspace{-1.8mm}Since the value in~(\ref{115b}) is positive, the large parenthesis in~(\ref{117}) is also positive and thus, the square mass in~(\ref{51}) is also positive. Using conditions~(\ref{108}), we can conclude that the square masses in~(\ref{50}),~(\ref{54b}),~(\ref{55b}) and~(\ref{58a}) are also positive quantities, with the hierarchy  $(M_{he_3}^2,M_{ho_1}^2,M_{h_2^\pm}^2)>M_{he_4}^2=M_{ho_2}^2$.

At the global minimum, the Higgs potential now becomes
\begin{equation}\label{117a}
V_{\textrm{min.}}=\frac{1}{4}\mu_1^2\;(v_1^2+V_1^2)+\frac{1}{4}\mu_3^2\;v_3^2<0.
\end{equation}
Therefore, in order to have the deepest minimum value for the potential for this particular vacuum structure, the following conditions are highly suggested
\begin{equation}
v_1,V_1,v_3\ne0.
\end{equation}

\subsection{\label{sec:sec53} The general case $v_1, V_1,v_2,V_2\neq0$}
In this case, we want the global minimum of potential be located at
$\tilde{\boldsymbol P}_5$. Looking at the Table~(\ref{c3}),  we choose this point as the global minimum, because for it $\langle\Phi_1\rangle$ and $\langle\Phi_2\rangle$ are LD and the VEV configuration presents in Eqs.  (\ref{1aa}) and (\ref{2aa}) are reproduced. For that, it is necessary that
\begin{equation}
\label{126}
\lambda_7+2\,\lambda_{10}<0.
\end{equation}
At the same time let us eliminate the possibility that the exceptional values $v_7$ and $w_8$ became global minima, which is reached  by making them negatives, that is 
\begin{equation}
\label{106a}
\lambda_8,\, \lambda_9>0,
\end{equation}
in such a way that the square mass in~(\ref{mh1pm}) becomes positive.

To exclude $\tilde{\boldsymbol P}_6$ as the global minimum it is sufficient to assume that $u_5>u_6$ which is achieved as far as 
\begin{equation}
\label{127}
\lambda_{10}<0.
\end{equation}
%
If we assume that
\begin{equation}
\label{108ab}
 \lambda_4<0,\quad \lambda_5<0,\quad\textrm{and}\quad\lambda_6<0,
\end{equation}
and taking into account~(\ref{71aa}),~(\ref{74aa}),~(\ref{76aa}),~(\ref{93a}) and~(\ref{126}), 
it implies that $u_1<0,w_2<0,u_3<0,w_4<0,v_9<0,u_{13}<0,v_{13}<0,u_{14}<0,w_{14}<0,v_{15}<0$ and $w_{15}<0$.  Finally,  the remaining Lagrange multipliers  $u_{10}<0,v_{10}<0,u_{11}<0,v_{11}<0,u_{12}<0$ and $u_{12}<0 $ are negative, when the corresponding points, are stationary points respectively. From conditions~(\ref{71aa}),~(\ref{76aa}),~(\ref{126}),~(\ref{106a}),~(\ref{127}) and~(\ref{108ab}),   the inequalities (\ref{20}),~(\ref{24}) and~(\ref{mh1pm}) are immediately satisfied.

As you can observe, being exhaustive in our reasoning,  the global minimum remains at  $ \tilde{\boldsymbol P}_5$, 
and it is given by
{\footnotesize
\begin{equation}\label{128}
\begin{pmatrix}
K_5= \frac{8\,\mu_3^2\,\left[\,w_{15}\,(u_5-u_3)\,+\,v_{15}\,(u_5-u_1)\right]}{d}
\\
L_5=\frac{8\,\mu_3^2\,[w_{15}\,(u_5-u_3)+u_5^2]-8\,\mu_1^2\,u_{13}\,w_4-2(2\lambda_4\mu_3^2-\lambda_5\mu_2^2-\lambda_6\mu_1^2)u_5}{d}
\\
M_5=\frac{8\,\mu_3^2\,[v_{15}\,(u_5-u_1)+u_5^2]-8\,\mu_1^2\,u_{13}\,w_4-2(2\lambda_4\mu_3^2-\lambda_5\,\mu_2^2-\lambda_6\,\mu_1^2)u_5}{d}
\\
\frac{16\,|\mu_3^2|\,\sqrt{(u_5-u_1)(u_5-u_3)\,v_{15}\,w_{15}}}{|d|}
\\
0\\
0\\
0\\
0\\
0
\end{pmatrix},
\end{equation}}
%
\hspace{-.8mm}with
\begin{equation*}
\begin{split}
d&=\lambda_1\,(4\lambda_2\,\lambda_3-\lambda_6^2)+\lambda_2\,(4\lambda_1\,\lambda_3-\lambda_5^2)
\\
&-\lambda_3\left[4\lambda_1\lambda_2+(\lambda_4-4\,u_5)^2\right]+\lambda_5\,\lambda_6\,(\lambda_4-4\,u_5),
\end{split}
\end{equation*}
where the first two terms of $d$ are positive and the last ones negative. In light of the    results above, we can  choose $\lambda_1,\lambda_2$ and $\lambda_3$ as larger as necessary such that 
\begin{equation}
 d>0,
\end{equation}
as can been observed from {\scriptsize $\lim_{\lambda_{1,2,3}\to{+}\infty}{d}\approx4\lambda_1\lambda_2\lambda_3=\mathcal{O}(\lambda^3)>0$}
 and similarly {\scriptsize $\lim_{\lambda_{1,2,3}\to{+}\infty}{L_5,M_5}=\mathcal{O}(1/\lambda)>0$}, such that it is possible to find cases for which
\begin{equation}
 L_5>0,\quad \textrm{and}\quad M_5>0.
\end{equation}
The fourth entry in~(\ref{128}) is taken positive by assuming positive VEV. Since we are in the domain given by~(\ref{100}), we must have
%
{\footnotesize
\begin{align}
\label{130}
 \tilde{\boldsymbol P}_5\cdot \tilde g_1\cdot \tilde{\boldsymbol P}_5&=0,
\\
\label{127b}
\begin{split}
\tilde{\boldsymbol P}_5\cdot \tilde g_2 \cdot \tilde{\boldsymbol P}_5&=
-64\,\mu_3^2\,w_{15}\,(u_5-u_3)\,\left[4\,\mu_1^2\,u_{13}\,w_4-4\,\mu_3^2\,u_5^2\right.
\\
&\left.-(\lambda_6\,\mu_1^2+\lambda_5\,\mu_2^2-2\lambda_4\,\mu_3^2)\,u_5\right]/d^2>0,
\end{split}
\\
\label{128b}
\begin{split}
\tilde{\boldsymbol P}_5\cdot \tilde g_3 \cdot \tilde{\boldsymbol P}_5&=
-64\,\mu_3^2\,v_{15}\,(u_5-u_1)\,\left[4\,\mu_1^2\,u_{13}\,w_4-4\,\mu_3^2\,u_5^2\right.
\\
&\left.-(\lambda_6\,\mu_1^2+\lambda_5\,\mu_2^2-2\lambda_4\,\mu_3^2)\,u_5\right]/d^2>0.
\end{split}
\end{align}}
where you can observe too that  {\footnotesize $\lim_{\lambda_{1,2,3}\to{+}\infty}\linebreak{\left(\tilde{\boldsymbol P}_5\cdot \tilde g_2 \cdot \tilde{\boldsymbol P}_5\:,\:\tilde{\boldsymbol P}_5\cdot \tilde g_3 \cdot \tilde{\boldsymbol P}_5\right)}=\mathcal{O}(1/\lambda^2)>0$}.

The expectation values satisfy also
{\scriptsize
\begin{align}
 &\langle\underline K\rangle=
\begin{pmatrix}
 \langle\Phi_1^\dag\Phi_1\rangle& \langle\Phi_2^\dag\Phi_1\rangle\\
\langle\Phi_1^\dag\Phi_2\rangle& \langle\Phi_2^\dag\Phi_2\rangle
\end{pmatrix}
\\\nonumber
&=
\begin{pmatrix}
 \frac{8\,\mu_3^2\,w_{15}\,(u_5-u_3)}{d}& \frac{8\,|\mu_3^2|\,\sqrt{(u_5-u_1)(u_5-u_3)\,v_{15}
\,w_{15}}}{|d|} \\
\frac{8\,|\mu_3^2|\,\sqrt{(u_5-u_1)(u_5-u_3)\,v_{15}\,w_{15}}}{|d|}&
\frac{8\,\mu_3^2\,v_{15}\,(u_5-u_1)}{d}
\end{pmatrix},
\\
\label{130b}
&\langle\underline L\rangle=
\begin{pmatrix}
 \langle\Phi_1^\dag\Phi_1\rangle& \langle\Phi_3^\dag\Phi_1\rangle\\
\langle\Phi_1^\dag\Phi_3\rangle& \langle\Phi_3^\dag\Phi_3\rangle
\end{pmatrix}
\\\nonumber
&=
\begin{pmatrix}
 \frac{8\,\mu_3^2\,w_{15}\,(u_5-u_3)}{d}& 0\\
0&-\frac{2\,[4\,\mu_1^2\,u_{13}\,w_4-4\,\mu_3^2\,u_5^2-(\lambda_6\,\mu_1^2+\lambda_5\,\mu_2^2-2\lambda_4\,\mu_3^2)\,u_5]}{d}
\end{pmatrix},
\\
&\langle\underline M\rangle=
\begin{pmatrix}
 \langle\Phi_2^\dag\Phi_2\rangle& \langle\Phi_3^\dag\Phi_2\rangle\\
\langle\Phi_2^\dag\Phi_3\rangle& \langle\Phi_3^\dag\Phi_3\rangle
\end{pmatrix}
\\
&=
\begin{pmatrix}
 \frac{8\,\mu_3^2\,v_{15}\,(u_5-u_1)}{d}& 0\\
0&-\frac{2\,[4\,\mu_1^2\,u_{13}\,w_4-4\,\mu_3^2\,u_5^2-(\lambda_6\,\mu_1^2+\lambda_5\,\mu_2^2-2\lambda_4\,\mu_3^2)\,u_5]}{d}
\end{pmatrix}. \nonumber
\end{align}}
%
\hspace{-1.5mm}This shows that $\langle\Phi_3\rangle$ is orthogonal to $\langle\Phi_1\rangle$ and~$\langle\Phi_2\rangle$, and at the same time $\langle\Phi_1\rangle$ and~$\langle\Phi_2\rangle$ are LD due to the relation (\ref{130}). It also shows that generality is not lost by taking positive VEV in (\ref{1aa})  and~(\ref{2aa}). We then have
{\footnotesize
\begin{align}
\label{134}
\frac{v_1^2+V_1^2}{2}&=\frac{8\,\mu_3^2\,w_{15}\,(u_5-u_3)}{d}, \\ \label{135}
\frac{v_2^2+V_2^2}{2}&=\frac{8\,\mu_3^2\,v_{15}\,(u_5-u_1)}{d}, \\ \label{136}
\begin{split}
\frac{v_3^2}{2}&=-\frac{8\,\mu_1^2\,u_{13}\,w_4-8\,\mu_3^2\,u_5^2}{d} \\ 
&+\frac{2\,(\lambda_6\,\mu_1^2+\lambda_5\,\mu_2^2-2\lambda_4\,\mu_3^2)\,u_5}{d}, 
\end{split}
\\ \label{137}
\frac{v_1\,v_2+V_1\,V_2}{2}&=\frac{8\,|\mu_3^2|\,\sqrt{(u_5-u_1)(u_5-u_3)\,v_{15} 
\,w_{15}}}{|d|},
\end{align}}
The solution to (\ref{134}),~(\ref{135}) and~(\ref{136}) coincide with the expressions given in~(\ref{16a}). Also, from the relations (\ref{ldvev}), (\ref{134}), (\ref{135}) and~(\ref{137}) we can see that the VEV satisfy the following relations 
\begin{equation}
\label{136a}
 \frac{v_1^2}{v_2^2}= \frac{V_1^2}{V_2^2}=\frac{w_{15}\,(u_5-u_3)}{v_{15}\,(u_5-u_1)}=\alpha^2,
\end{equation}
where 
\begin{equation}\label{alpha}
\alpha=\sqrt{\frac{w_{15}\,(u_5-u_3)}{v_{15}\,(u_5-u_1)}}
\end{equation}
is the proportionality factor between $\langle\Phi_1\rangle$ and $\langle\Phi_2\rangle$ as stated in (\ref{ld3}), which allow us to connect the value of $\alpha$ to the Lagrange multipliers.

At the global minimum the Higgs potential becomes
\begin{equation}
V_{\textrm{min.}}=\frac{1}{4}\mu_1^2\,(v_1^2+V_1^2)+\frac{1}{4}\mu_2^2\,(v_2^2+V_2^2)+\frac{1}{4}\mu_3^2\,v_3^2<0,
\end{equation}
which reproduces Eq.~(\ref{117a}) in the limit $v_2=V_2=0$. Therefore, in order to have the deepest minimum value for the potential for this particular vacuum structure, the following conditions are highly suggested
\begin{align}
\label{123}
v_1,V_1,v_2,V_2,v_3\ne0.
\end{align}

\section{\label{sec:sec6} Conclusions}
In this paper we have presented original results related to the scalar sector of some 3-3-1 models without exotic electric charges. An exhaustive study of the scalar potential with 3 scalar triplets $\Phi_1,\;\;\Phi_2$ and $\Phi_3$ and VEV as introduced in Sects.~(\ref{sec:sec231}) and (\ref{sec:sec232}); potential which does not include the possible cubic term  according to the discrete symmetry $\Phi_1\rightarrow-\Phi_1$ imposed, has been carried through. This problem partially analyzed in the literature~\cite{vl,folo,bb3_1} had not been studied in a systematic way. 

In concrete we have:
\begin{itemize}
\item Looked for a consistent implementation of the Higgs mechanism.
\item Implemented a consistent electroweak symmetry breaking pattern.
\item Established the strong stability conditions for the scalar potential.
\item Found the stationary points of the scalar potential, except the ones coming from specific relations among the parameters $\mu_1^2$, $\mu_2^2$ and $\mu_3^2$ that we assume are not satisfied, in general.
\end{itemize}

One outstanding new result is that $\langle\Phi_1\rangle$ and $\langle\Phi_2\rangle$, the VEV of the two Higgs scalars with identical quantum numbers, must be proportional to each other, a necessary condition in order to properly implement the Higgs mechanism, and achieve a consistent electroweak symmetry breaking; besides, the proportionality constant is connected with the Lagrange multipliers (which in turn are connected to the other parameters of the potential) via Eq.~(\ref{alpha}).

Other important result is that, from the nine possible vacuum structures compatible with the stated LD constraint, only two are independent. Our analysis has been done for both structures.

But probably, the most important conclusion of our study is the existence~(as a sufficient condition), in the scalar potential, of a global minimum stationary point for each one of the two cases of VEV considered, being it, at the same time, compatible with the stability conditions imposed in the strong sense.  Stability conditions and the global minimum point were found via the orbit gauge method, implemented in this case for three scalar triplets.  Although the method can give unphysical VEVs.

\section*{Acknowledgments}
We both acknowledge the warm hospitality from the ``Laboratorio de F\'isica Te\'orica de la Universidad de La Plata, in La Plata, Argentina'', during several stages of the development of the present work. Also, very illuminating were several conversations with professors C. Garc\'ia-Canal, D. G\'omez-Dumm and H. Fanchiotti. We thank in special C. Garc\'ia-Canal for a critical reading of the original manuscript. Y.G. acknowledges financial support and leave of absence from ``Universidad de Nari\~no'' in Pasto, Colombia. The present work is part of the doctoral thesis of Y.G.
\appendix

\section{A review of the algebraic method used}
\label{aA}
Let us briefly review in this appendix, and following Refs. \cite{b1} and \cite{b2}, a new algebraic approach used to determine the global minimum of the Higgs scalar potential, its stability, and the spontaneous symmetry breaking from $SU(2)_L\otimes U(1)_Y$ down to $U(1)_{EM}$, in the extension of the SM known as the THDM. This mathematical approach and its notation was used in Secs~(\ref{sec:sec4}) and~(\ref{sec:sec5}) of this paper, generalizing it to the study of the scalar sector of some 3-3-1 models.

Stability, and the stationary points of the scalar potential for the THDM can be analyzed in terms of four real quantities given by
\begin{equation}\label{k0}
K_0=\sum_{i=1,2}\varphi_i^\dag\varphi_i,\quad K_a=\sum_{i,j=1,2}(\varphi_i^\dag\varphi_j)\sigma^a_{ij},\quad (a=1,2,3).
\end{equation}
where $\varphi_1$ and $\varphi_2$ stand for two Higgs scalar field doublets with identical quantum numbers and 
$\sigma^a(a=1,2,3)$ are the Pauli spin matrices. The four vector $(K_0,\boldsymbol K)$ must lie on or inside the forward \textit{light cone}, that is
\begin{equation}
\label{19}
 K_0\geq 0,\quad K_0^2-\boldsymbol K^2\geq 0.
\end{equation}
Then the positive and hermitian $2\times 2$ matrix
\begin{equation}
\label{6aa}
\underline K=  
\begin{pmatrix}
             \varphi_1^\dag\varphi_1 & \varphi_2^\dag\varphi_1\\
\varphi_1^\dag\varphi_2 & \varphi_2^\dag\varphi_2
              \end{pmatrix}
\end{equation}
may be written as 
\begin{equation}\label{6ab}
\underline K_{ij}=\frac{1}{2}(K_0\delta_{ij}+K_a\sigma^a_{ij}).
\end{equation}
Inverting Eq.~(\ref{k0}) one obtains
\begin{equation}
\begin{split}
 \varphi_1^\dag\varphi_1=(K_0+K_3)/2,\quad  \varphi_1^\dag\varphi_2=(K_1+iK_2)/2,\\
 \varphi_2^\dag\varphi_2=(K_0-K_3)/2,\quad  \varphi_2^\dag\varphi_1=(K_1-iK_2)/2\:.
\end{split}
\end{equation}
The most general $SU(2)_L\otimes U(1)_Y$ invariant Higgs scalar potential can thus be expressed as
\begin{subequations}
\label{10aa}
\begin{align}
 V(\varphi_1,\varphi_2)&=V_2+V_4;\\
V_2&=\xi_0K_0+\xi_aK_a,\\
V_4&=\eta_{00}K_0^2+2K_0\eta_aK_a+K_a\eta_{ab}K_b,
\end{align}
\end{subequations}
where the 14 independent parameters $\xi_0,\;\xi_a,\; \eta_{00},\; \eta_{a}$ and $\eta_{ab}=\eta_{ba}$ are real. Subsequently, it is defined $\boldsymbol K=(K_a),\; \boldsymbol{\xi}=(\xi_a),\; \boldsymbol\eta=(\eta_a)$ and $E=(\eta_{ab})$.

\subsection{\label{sec:seca1}Stability}
From (\ref{10aa}), for~\mbox{$K_0 > 0$} and defining $\boldsymbol{k} = \boldsymbol{K} / K_0$, one obtains
\begin{align}
\label{eq-vk}
V_2 &= K_0\, J_2(\boldsymbol{k}),&
J_2(\boldsymbol{k}) &:= \xi_0 + \boldsymbol{\xi}^\mathrm{T} \boldsymbol{k},\\
\label{eq-vk4}
V_4 &= K_0^2\, J_4(\boldsymbol{k}),&
J_4(\boldsymbol{k}) &:= \eta_{00} 
  + 2 \boldsymbol{\eta}^\mathrm{T} \boldsymbol{k} + \boldsymbol{k}^\mathrm{T} E \boldsymbol{k},
\end{align}
where the functions $J_2(\boldsymbol{k})$ and $J_4(\boldsymbol{k})$
on the domain \mbox{$|\boldsymbol{k}| \leq 1$} have been introduced.

For the potential to be stable, it must be bounded from below.
The stability is determined by the behavior of $V=V_1+V_2$ in the limit
\mbox{$K_0 \rightarrow \infty$}, hence by the signs of
\mbox{$J_2(\boldsymbol{k})$} and/or \mbox{$J_4(\boldsymbol{k})$} in~(\ref{eq-vk}) and~(\ref{eq-vk4}).
In the analysis, only the \emph{strong} criterion for stability is analyzed. That is, the stability is going to be determined solely by $V_4$ (the $V$ quartic term). That is, we demand that
\begin{equation}
\label{eq-jinq}
J_4(\boldsymbol{k}) > 0 \quad \text{for all }\left\lvert {\boldsymbol{k}}\right\rvert \leq 1.
\end{equation}

To assure that $J_4(\boldsymbol{k})$ is always positive, it is sufficient to
consider its value for all its stationary points
on the domain \mbox{$\left\lvert {\boldsymbol{k}}\right\rvert < 1$} and for all the stationary points on the
boundary~\mbox{$|\boldsymbol{k}| = 1$}. 
This leads to bounds on the parameters $\eta_{00}$, $\eta_{a}$ and $\eta_{ab}$ of the quartic term~$V_4$ of the scalar potential. The regular solutions are obtained by inverting the matrix $(E-u)$ 
which appears in the calculations, where $u$ is the Lagrange multiplier associated to the boundary condition under consideration. When the matrix is not invertible, we say that we have an exceptional solution.
The two cases \mbox{$\left\lvert {\boldsymbol{k}}\right\rvert < 1$} and
\mbox{$\left\lvert {\boldsymbol{k}}\right\rvert = 1$} lead to the functions
\begin{align}
\label{eq-flam}
f(u) & =  u + \eta_{00} - \boldsymbol{\eta}^\mathrm{T} (E - u)^{-1} \boldsymbol{\eta},\\
\label{eq-flampr}
f'(u) & =  1 - \boldsymbol{\eta}^\mathrm{T} (E - u)^{-2} \boldsymbol{\eta},
\end{align}
so that for all ``regular'' stationary points~$\boldsymbol{k}$
of~$J_4(\boldsymbol{k})$
\begin{align}
\label{11d}
f(u) &= \left. J_4(\boldsymbol{k}) \right|_{\substack{stat}},\quad \mbox{and}\\
f'(u) &= 1 - \boldsymbol{k}^2
\end{align}
hold, where $u=0$ must be set for the solution with $\left\lvert {\boldsymbol{k}}\right\rvert<1$.
There are stationary points of~$J_4(\boldsymbol{k})$ with $\left\lvert {\boldsymbol{k}}\right\rvert<1$
and $\left\lvert {\boldsymbol{k}}\right\rvert=1$ exactly if $f'(0)>0$ and $f'(u)=0$, respectively,
and the value of $J_4(\boldsymbol{k})$ is then given by $f(u)$.

In a basis where \mbox{$E = {\rm diag}(\mu_1, \mu_2, \mu_3)$} it is obtained
\begin{align}
\label{eq-fdiag}
f(u) &= u + \eta_{00} - \sum_{a = 1}^3 \frac{\eta_a^2}{\mu_a - u},\\
\label{eq-fprd}
f'(u) &= 1 - \sum_{a = 1}^3 \frac{\eta_a^2}{(\mu_a - u)^2}.
\end{align}
The derivative~\mbox{$f'(u)$} has at most 6 zeros.
Notice that there are no exceptional solutions
if in this basis all three components of~$\boldsymbol{\eta}$ are different from zero.

Consider now the functions $f(u)$ and $f'(u)$ and denote by $I$
\begin{equation}
\label{eq-idef}
I = \{ u_1, \dots, u_n \},
\end{equation}
the set of values $u_j$ for which $f'(u_j)=0$. Add $u_k=0$ to $I$ if $f'(0)>0$.
Consider then the eigenvalues $\mu_a$ ($a=1,2,3$) of $E$.
Add those $\mu_a$ to $I$ where $f(\mu_a)$ is finite and $f'(\mu_a) \geq 0$.
Then $n \leq 10$. The values of the function
$J_4(\boldsymbol{k})$
at its stationary points are given by
\begin{equation}
\left. J_4(\boldsymbol{k}) \right|_{\substack{stat}} = f(u_i)
\end{equation}
with $u_i \in I$. In Ref.~\cite{yith} it was shown that the stationary point in $I$ having the smallest value, will produce the smallest value of $J_4(\boldsymbol k)$ in the domain $|\boldsymbol k|\leq1$. Then, one states the following theorem
\begin{theorem}
\label{t1}
The global minimum of the function $J_4(\boldsymbol k)$, in the domain $|\boldsymbol k|\leq1$,  is given and guaranteed by the stationary point of the set $I$ with the smallest value.
\end{theorem}
Namely, this result guarantees strong stability if $f(u)>0$, where $u$ is the smallest value of $I$.
The potential is unstable if we have $f(u)<0$. If $f(u)=0$ we have to consider in addition $J_2(\boldsymbol k)$ in order to decide on the stability of the potential. 

\subsection{\label{sec:seca2}Stationary points}
After the stability analysis is done, the next step is to determine the location of the stationary points of the scalar potential, since among these points the local and global minima are found. To this end it is defined
\begin{equation}\label{tildepar}
\boldsymbol{\tilde{K}} = \begin{pmatrix} K_0\\ \boldsymbol{K} \end{pmatrix},\quad
\boldsymbol{\tilde{\xi}}  = \begin{pmatrix} \xi_0\\ \boldsymbol{\xi} \end{pmatrix},\quad
\tilde{E} = \begin{pmatrix} \eta_{00} & \boldsymbol{\eta}^\mathrm{T}\\
                           \boldsymbol{\eta} & E \end{pmatrix}.
\end{equation}
In this notation the potential~(\ref{10aa}) reads
\begin{equation}
\label{eq-vtil}
V =\boldsymbol{\tilde{K}}^\mathrm{T}\boldsymbol{\tilde{\xi}}  + \boldsymbol{\tilde{K}}^\mathrm{T}
\tilde{E} \boldsymbol{\tilde{K}}
\end{equation}
and is defined on the domain
\begin{equation}
\label{eq-domv}
\boldsymbol{\tilde{K}}^\mathrm{T} \tilde{g} \boldsymbol{\tilde{K}}\geq 0,
\qquad K_0 \ge 0,
\end{equation}
with
\begin{equation}
\tilde{g} = \begin{pmatrix} 1 & \phantom{-}0 \\ 0 & -\mathbbm{1} \end{pmatrix}.
\end{equation}
For the discussion of the stationary points of~$V$, three different cases must be distinguished:
$\boldsymbol{\tilde{K}}=0$,
$K_0 > \left\lvert {\boldsymbol{K}} \right\rvert$ which are the solutions inside the forward light cone, 
and
$K_0 = \left\lvert {\boldsymbol{K}} \right\rvert > 0$ which are the solutions on the forward light cone.

The trivial configuration $\boldsymbol{\tilde{K}}=0$
is a stationary point of the potential with~$V=0$,
as a direct consequence of the definitions.
The stationary points of~$V$ in the inner part of the domain,
$K_0>\left\lvert {\boldsymbol{K}} \right\rvert$,
are given by
\begin{equation}
\label{eq-statin}
\tilde{E} \boldsymbol{\tilde{K}}  = - \frac{1}{2} \boldsymbol{\tilde{\xi}},
\quad
\text{with}
\quad
\boldsymbol{\tilde{K}}^\mathrm{T} \tilde{g} \boldsymbol{\tilde{K}}>0
\quad
\text{and}
\quad
K_0>0.
\end{equation}
The stationary points of $V$ on the domain boundary
$K_0 = \vert \boldsymbol{K}\vert > 0$ are stationary points of the function
\begin{equation}
\tilde{F}\big(\boldsymbol{\tilde{K}}, w \big) = V - w \boldsymbol{\tilde{K}}^\mathrm{T}
\tilde{g} \boldsymbol{\tilde{K}},
\end{equation}
where $w$ is a Lagrange multiplier. The relevant stationary points of
\mbox{$\tilde{F}$} are given by
\begin{equation}
\label{eq-stap}
\big(\tilde{E} -w \tilde{g} \big) \boldsymbol{\tilde{K}} =
-\frac{1}{2} \boldsymbol{\tilde{\xi}},
\quad \text{with}\quad \boldsymbol{\tilde{K}}^\mathrm{T} \tilde{g} \boldsymbol{\tilde{K}}=0 \quad \text{and}\quad K_0 >0.
\end{equation}
For \emph{any} stationary point, the potential is given by
\begin{equation}
\label{eq-statexpl}
V|_{\substack{stat}} = \frac{1}{2} \boldsymbol{\tilde{K}}^\mathrm{T} \boldsymbol{\tilde{\xi}}= -
\boldsymbol{\tilde{K}}^\mathrm{T} \tilde{E} \boldsymbol{\tilde{K}}.
\end{equation}
Similarly to the stability analysis in Sec.~\ref{sec:seca1}, a unified description for the regular stationary points of~$V$ with
$K_0>0$ for both \mbox{$|\boldsymbol{K}| < K_0$} and \mbox{$|\boldsymbol{K}| = K_0$} can be used by defining the functions
\begin{align}
\label{eq-ftil}
\tilde{f}(w) &= -\frac{1}{4} \boldsymbol{\tilde{\xi}}^{\, \rm T}
 \big( \tilde{E} - w \tilde{g} \big)^{-1} \boldsymbol{\tilde{\xi}},\\
\label{eq-ftilpr}
\tilde{f}'(w) &= - \frac{1}{4} \boldsymbol{\tilde{\xi}}^{\, \rm T}
\big( \tilde{E}-w\tilde{g} \big)^{-1} \tilde{g}
  \big( \tilde{E}- w \tilde{g} \big)^{-1} \boldsymbol{\tilde{\xi}}.
\end{align}
Denoting the first component of \mbox{$\boldsymbol{\tilde{K}}(w)$} as $K_0(w)$
the following theorem holds
\begin{theorem}
\begin{samepage}
\label{classes-statpoints}
The stationary points of the potential are given by
\begin{itemize}
\item[(I\,a)]
$\boldsymbol{\tilde{K}} = \boldsymbol{\tilde{K}}(0)$
if $\tilde{f}'(0) < 0$,\  $K_0(0)>0$ and \mbox{$\det \tilde{E} \neq 0$},
\item[(I\,b)]
solutions $\boldsymbol{\tilde{K}}$ of~\eqref{eq-statin}
if \mbox{$\det \tilde{E} = 0$},
\item[(II\,a)]
$\boldsymbol{\tilde{K}}=\boldsymbol{\tilde{K}}(w)$
for $w$ with \mbox{$\det (\tilde{E} - w \tilde{g}) \neq 0$},\ 
 \mbox{$\tilde{f}'(w) = 0$} and
 $K_0(w)>0$,
\item[(II\,b)]
solutions $\boldsymbol{\tilde{K}}$ of~\eqref{eq-stap}
for $w$ with \mbox{$\det (\tilde{E} - w\tilde{g}) = 0$},
\item[(III)]
$\boldsymbol{\tilde{K}}$ = 0.
\end{itemize}
\end{samepage}
\end{theorem}
In what follows it is assumed that the potential is stable.
For parameters fulfilling \mbox{$\xi_0 \geq |\boldsymbol{\xi}|$},
this immediately implies 
\mbox{$J_2(\boldsymbol{k})\ge 0$} and hence, from the strong condition~\eqref{eq-jinq}, 
$V>0$ for all \mbox{$\boldsymbol{\tilde{K}} \neq 0$}.
Therefore for these parameters the global minimum is at~$\boldsymbol{\tilde{K}}=0$.
This leads to the requirement
\begin{equation}
\label{eq-cond}
\xi_0 < |\boldsymbol{\xi}|.
\end{equation}
Also, it is obtained
\begin{equation}
\label{28aa}
\left. \frac{\partial V}{\partial K_0} \right|_{
  \begin{subarray}{l} \boldsymbol{k}\;\text{fixed},\\
                    K_0 = 0 \end{subarray}
} = \xi_0 +  \boldsymbol{\xi}^\mathrm{T} \boldsymbol{k}
  < 0 
\end{equation} 
for some $\boldsymbol{k}$, i.e.\ the global minimum of $V$ lies
at~\mbox{$\boldsymbol{\tilde{K}} \neq 0$} with
\begin{equation}
\label{29b}
 V|_{\substack{min}} <0.
\end{equation}
Firstly, consider $p_0 = \left\lvert {\boldsymbol{p}} \right\rvert$.
From~\eqref{eq-vtil} and \eqref{eq-stap} it follows
\begin{equation}
\label{55a}
\left. \frac{\partial V}{\partial K_0} \right|_{
  \begin{subarray}{l} \boldsymbol{K}\;\text{fixed},\\
                      \boldsymbol{\tilde{K}}=\boldsymbol{\tilde{p}} \end{subarray}
}
 = \xi_0 + 2 (\tilde{E}\,\boldsymbol{\tilde{p}})_0
 = 2 w_p\, p_0.
\end{equation}
If $w_p<0$, there are points $\boldsymbol{\tilde{K}}$ with $K_0>p_0$,
$\boldsymbol{K}=\boldsymbol{p}$ and lower potential in the neighborhood of
$\boldsymbol{\tilde{p}}$, which therefore cannot be a minimum.
The conclusion is that in a theory with the required electroweak symmetry breaking~(EWSB) the global minimum must have
a Lagrange multiplier such that $w_0 \geq 0$, and for the THDM, the global minimum lies on the stationary points of the classes $(IIa)$ and $(IIb)$ of theorem 2, with the largest Lagrange multiplier~\cite{b1}.

\section{ Linear dependence between $\langle\Phi_1\rangle$ and $\langle\Phi_2\rangle$}
\label{aB}
In this appendix we study the consequences of a linear dependence between $\langle\Phi_1\rangle$ and $\langle\Phi_2\rangle$.

As in the main text we use the VEV 
\[ \langle\Phi_1\rangle=
\frac{1}{\sqrt{2}}
\begin{pmatrix}
 0\\
v_1\\
V_1
\end{pmatrix},\quad
\langle\Phi_2\rangle=
\frac{1}{\sqrt{2}}
\begin{pmatrix}
 0\\
v_2\\
V_2
\end{pmatrix},\]
\[\langle\Phi_3\rangle=
\frac{1}{\sqrt{2}}
\begin{pmatrix}
v_3\\
0\\
0
\end{pmatrix}.\]

The LD between $\langle\Phi_1\rangle$ and $\langle\Phi_2\rangle$ can be written as
\begin{equation}\label{ld3}
\langle\Phi_1\rangle=\alpha \langle\Phi_2\rangle,
\end{equation}
where $\alpha$ is a constant. Eq.~(\ref{ld3}) implies that $v_1=\alpha v_2$ and $V_1=\alpha V_2$, which combine to produce the constraint $v_2V_1=v_1V_2$.

Now, the nine $U(3)$ generators are
%
{\small
\begin{equation}\label{u3gen}
 \begin{split}
  I_3&=\sqrt{\frac{2}{3}}\begin{pmatrix}1 & 0 & 0\cr 0 & 1 & 0\cr 0 & 0 & 1\end{pmatrix},\:
\lambda_1=\begin{pmatrix}0 & 1 & 0\cr 1& 0& 0\cr 0 & 0 & 0\end{pmatrix},\:
\lambda_2=\begin{pmatrix}0 & -i & 0\cr i& 0& 0\cr 0 & 0 & 0\end{pmatrix},\:
\\
\lambda_3&=\begin{pmatrix}1 & 0 & 0\cr 0& -1& 0\cr 0 & 0 & 0\end{pmatrix},\:
\lambda_4=\begin{pmatrix}0 & 0 & 1\cr 0& 0& 0\cr 1 & 0 & 0\end{pmatrix},\:
\lambda_5=\begin{pmatrix}0 & 0 & -i\cr 0& 0& 0\cr i & 0 & 0\end{pmatrix},
\\
\lambda_6&=\begin{pmatrix}0 & 0 & 0\cr 0& 0& 1\cr 0 & 1 & 0\end{pmatrix},\:
\lambda_7=\begin{pmatrix}0 & 0 & 0\cr 0& 0& -i\cr 0 & i & 0\end{pmatrix},\:
\lambda_8=\frac{1}{\sqrt{3}}\begin{pmatrix}1 & 0 & 0\cr 0& 1& 0\cr 0 & 0& -2\end{pmatrix},
 \end{split}
\end{equation}}
where $\lambda_i(i=1,\ldots8)$ are the eight Gell-Mann unitary matrices for $SU(3)$.

Let us now show that the LD in Eq.~(\ref{ld3}) with the additional constraint $\langle\Phi_3\rangle\ne0$, implies that either $I_3$, or a linear combination of the generators in (\ref{u3gen}) which includes $I_3$, remains unbroken, with the consequence that the appearance of an extra zero mass Goldstone Bosons is avoided.

The algebra shows that the most general new unbroken generator is given by the following linear combination:

\begin{equation}\label{unnw}
G=aI_3+b\lambda_3+c\lambda_8+d\lambda_6=
\begin{pmatrix}0 & 0 & 0\cr 0 & {V_1}^{2} & -v_1\,V_1\cr 0 & -v_1\,V_1 & {v_1}^{2}\end{pmatrix},
\end{equation}
where
\begin{eqnarray*}
a=&\frac{V_1^2+v_1^2}{2},\quad b=-\frac{V_1^2}{2},\\
c=&\frac{{V_1}^{2}-2\,{v_1}^{2}}{2\,\sqrt{3}},\quad d= -v_1\,V_1.
\end{eqnarray*}

That $G$ remains unbroken can be seen by the fact that $G \langle\Phi_1\rangle=0$ by direct calculation, $G \langle\Phi_2\rangle=0$ is a consequence of the relation (\ref{ldvev}), and $G \langle\Phi_3\rangle=0$ is trivial.

Since $Tr.G=v_1^2+V_1^2\neq 0$, the new unbroken generator in Eq.~(\ref{unnw}) is such that $G\in U(3)$  but $G\notin SU(3)$.


\section{ Discrete symmetry in the scalar potential}
\label{aC}
Under assumption of the discrete symmetry $\Phi_1\rightarrow -\Phi_1$,
the most general potential obtained from (\ref{7a}), can then be written in the following
form:
%
{\footnotesize
\begin{equation}\label{B1}
\begin{split}
 V&(\Phi_1,\Phi_2,\Phi_3)=\mu_1^2\Phi_1^\dag\Phi_1+\mu_2^2\Phi_2^\dag\Phi_2+\mu_3^2\Phi_3^\dag\Phi_3
+\lambda_1(\Phi_1^\dag\Phi_1)^2
\\
&+\lambda_2(\Phi_2^\dag\Phi_2)^2+\lambda_3(\Phi_3^\dag\Phi_3)^2
+\lambda_4(\Phi_1^\dag\Phi_1)(\Phi_2^\dag\Phi_2)
\\
&+\lambda_5(\Phi_1^\dag\Phi_1)(\Phi_3^\dag\Phi_3)
+\lambda_6(\Phi_2^\dag\Phi_2)(\Phi_3^\dag\Phi_3)
+\lambda_7(\Phi_1^\dag\Phi_2)(\Phi_2^\dag\Phi_1)\\
&+\lambda_8(\Phi_1^\dag\Phi_3)(\Phi_3^\dag\Phi_1)
+\lambda_9(\Phi_2^\dag\Phi_3)(\Phi_3^\dag\Phi_2)+
(f\:\Phi_1^\dag\Phi_2+f^*\:\Phi_2^\dag\Phi_1)^2.
\end{split}
\end{equation}}
%
\hspace{-2mm}where the complex value $f$ is going to be used as $f=f_1+if_2$, with $f_j,\; j=1,2$ are two real parameters. With the new definitions of the scalar fields introduced in (\ref{newfields}), and by demanding that the VEV in~(\ref{1aa}) and~(\ref{2aa}) became stationary points of the potential, the following nine constraints must be satisfied
%
{\scriptsize
\begin{align}\label{B2}
\begin{split}
&\left.\frac{\partial V}{\partial H_1}\right|_{\textrm{fields}=0}=\\
&\left[2\,\mu_1^2 v_1+\lambda_4\,v_1\,{V_2}^{2}+\left( \lambda_7+4\,{f_1}^{2}\right) \,v_2\,V_1\,V_2+2\,\lambda_1\,v_1\,{V_1}^{2}\right.
\\
&+\left.\left( \lambda_7+\lambda_4+4\,{f_1}^{2}\right) \,v_1\,{v_2}^{2}+2\,\lambda_1\,{v_1}^{3}+\lambda_5\,{v_3}^{2}\,v_1\right]\Big{
/}2=0,
\end{split}
\\ \label{B3}
\begin{split}
&\left.\frac{\partial V}{\partial H^\prime_1}\right|_{\textrm{fields}=0}=\\
&\left[2\,\mu_1^2V_1+\left( \lambda_7+\lambda_4+4\,{f_1}^{2}\right) \,V_1\,{V_2}^{2}+\left( \lambda_7+4\,{f_1}^{2}\right) \,v_1\,v_2\,V_2\right.
\\
&+\left.2\,\lambda_1\,{V_1}^{3}+\left( \lambda_4\,{v_2}^{2}+2\,\lambda_1\,{v_1}^{2}+\lambda_5\,{v_3}^{2}\right) \,V_1\right]\Big{
/}2=0,
\end{split}
\\ \label{B4}
\begin{split}
&\left.\frac{\partial V}{\partial H_2}\right|_{\textrm{fields}=0}=\\
&\left\{2\,\mu_2^2v_2+2\,\lambda_2\,v_2\,{V_2}^{2}+\left( \lambda_7+4\,{f_1}^{2}\right) \,v_1\,V_1\,V_2+\lambda_4\,v_2\,{V_1}^{2}\right.
\\
&+\left.2\,\lambda_2\,{v_2}^{3}+\left[ \left( \lambda_7+\lambda_4+4\,{f_1}^{2}\right) \,{v_1}^{2}+\lambda_6\,{v_3}^{2}\right] \,v_2\right\}\Big{
/}2=0,
\end{split}
\\ \label{B5}
\begin{split}
&\left.\frac{\partial V}{\partial H^\prime_2}\right|_{\textrm{fields}=0}=\\
&\left\{2\mu_2^2V_2+2\lambda_2{V_2}^{3}+\left( \lambda_7+4\,{f_1}^{2}\right) \,v_1\,v_2\,V_1\right.
\\
&+\left.\left[ \left( \lambda_7+\lambda_4+4{f_1}^{2}\right) {V_1}^{2}+2\lambda_2{v_2}^{2}+\lambda_4{v_1}^{2}+\lambda_6{v_3}^{2}\right]V_2\right\}\Big{
/}2=0,
\end{split}
\\ \label{B6}
&\left.\frac{\partial V}{\partial H_3}\right|_{\textrm{fields}=0}\!\!\!\!\!\!\!=
\frac{v_3\left(
2\mu^2_3+ \lambda_6{V_2}^{2}+\lambda_5{V_1}^{2}+\lambda_6{v_2}^{2}+\lambda_5{v_1}^{2}+2\lambda_3{v_3}^{2}\right) }{2}\!\!=0,
\end{align}}
\begin{align}\label{B7}
&\left.\frac{\partial V}{\partial A_1}\right|_{\textrm{fields}=0}=
2f_1f_2\,v_2\,\left( V_1\,V_2+v_1\,v_2\right)=0,
\\
\label{B8}
&\left.\frac{\partial V}{\partial A_1^\prime}\right|_{\textrm{fields}=0}=
2f_1f_2\,V_2\,\left( V_1\,V_2+v_1\,v_2\right)=0,
\\
\label{B9}
&\left.\frac{\partial V}{\partial A_2}\right|_{\textrm{fields}=0}=
-2f_1f_2\,v_1\,\left( V_1\,V_2+v_1\,v_2\right)=0,
\\
\label{B10}
&\left.\frac{\partial V}{\partial A_2^\prime}\right|_{\textrm{fields}=0}=
-2f_1f_2\,V_1\,\left( V_1\,V_2+v_1\,v_2\right)=0.
\end{align}
A simple algebra shows that both operations [$v_1\times(\ref{B3})-V_1\times(\ref{B2})$] and   [$V_2\times(\ref{B4})-v_2\times(\ref{B5})$] produce the same relation
\begin{equation}\label{ac11}
( \lambda_7+4f_1^{2})( v_1\,V_2-v_2\,V_1)\,( V_1\,V_2+v_1\,v_2)=0,
\end{equation}
which must be satisfied in order to have a consistent set of equations~(\ref{B2})-(\ref{B5}). 

The two possible solutions to (\ref{ac11}) are $(v_1V_2-v_2V_1)=0$ and/or $(V_1V_2+v_1v_2)=0$. Obviously, (\ref{ac11}) is satisfied if $\langle\Phi_1\rangle$ and $\langle\Phi_2\rangle$ are LD. (For the unphysical case $\langle\Phi_3\rangle=0$ with $\langle\Phi_1\rangle$ and $\langle\Phi_2\rangle$ being linearly independent, the mathematical solution $V_1V_2=-v_1v_2$ is still available.)

But at the same time, the relations~(\ref{B7})-(\ref{B10}) must be satisfied, the alternative which remains for the physical case is that either the real or the imaginary part of $f$ become zero, that is
\begin{equation}
f_1=0\;\;\;\textrm{or}\;\;\; f_2=0,
\end{equation}
meaning that $f$ represents only one parameter, something which allow us to introduce the usual notation 
$|f|^2=\frac{\lambda_{10}}{2}$, with $\lambda_{10}$ either positive or negative.

\end{document}